\documentclass[final,3p,times]{elsarticle}

\usepackage[utf8]{inputenc}
\usepackage{mathtools}
\usepackage{amssymb}
\usepackage{amsmath}
\usepackage{graphicx}
\usepackage{booktabs}
\usepackage{multirow}
\usepackage{float}
\usepackage{subcaption}
\usepackage{bm}
\usepackage{color}
\usepackage{algorithm}
\usepackage{algpseudocode}
\usepackage{url}
\usepackage{hyperref}
\usepackage{adjustbox}
\usepackage{longtable}
\usepackage{stackengine}
\usepackage{dsfont}

\usepackage{tikz}

\newcommand*\circled[1]{\tikz[baseline=(char.base)]{
            \node[shape=circle,draw,inner sep=1pt] (char) {#1};}}
\stackMath

\journal{
Transportation Research Part A: Policy and Practice}

\begin{document}

\begin{frontmatter}

%% Title, Authors, and Affiliations
\title{A Short-Term Predict-Then-Cluster Framework for Meal Delivery Services}

\author[inst1]{Jingyi Cheng}
\author[inst1]{Shadi Sharif Azadeh\corref{cor1}}
\cortext[cor1]{Corresponding author}
\ead{s.sharifazadeh@tudelft.nl}

\affiliation[inst1]{
    organization={Delft University of Technology}, 
    addressline={Mekelweg 5}, 
    city={Delft},
    postcode={2628 CD},
    country={The Netherlands}
}

\textit{This manuscript is currently under review at Transportation Research Part A: Policy and Practice.}

%% Abstract
\begin{abstract}

Micro-delivery services offer promising solutions for on-demand city logistics, but their success relies on efficient real-time delivery operations and fleet management. 
On-demand meal delivery platforms seek to optimize real-time operations based on anticipatory insights into citywide demand distributions.    
To address these needs, this study proposes a short-term predict-then-cluster framework for on-demand meal delivery services. The framework utilizes ensemble-learning methods for point and distributional forecasting with multivariate features, including lagged-dependent inputs to capture demand dynamics. 
We introduce Constrained K-Means Clustering (CKMC) and Contiguity Constrained Hierarchical Clustering \textcolor{black}{with Iterative Constraint Enforcement (CCHC-ICE)} to generate dynamic clusters based on predicted demand and geographical proximity\textcolor{black}{, tailored to user-defined operational constraints.}
Evaluations of European and Taiwanese case studies demonstrate that the proposed methods outperform traditional time series approaches in both accuracy and computational efficiency. Clustering results demonstrate that the incorporation of distributional predictions effectively addresses demand uncertainties, improving the quality of operational insights. \textcolor{black}{Additionally, a simulation study demonstrates the practical value of short-term demand predictions for proactive strategies, such as idle fleet rebalancing, significantly enhancing delivery efficiency.}
By addressing demand uncertainties and operational constraints, our predict-then-cluster framework provides actionable insights for optimizing real-time operations. The approach is adaptable to other on-demand platform-based city logistics and passenger mobility services, promoting sustainable and efficient urban operations.

\end{abstract}

%% Keywords

\end{frontmatter}

%% Main Content Sections
\section{Introduction}
\label{sec:introduction}
``Click and pay, soon food is at the doorway" has become an urban lifestyle nowadays, driving the growth of the meal delivery industry into a global market exceeding 150 billion dollars \cite{McKinsey}.  The COVID-19 pandemic further accelerated demand due to restaurant closures and government-imposed restrictions. In response, more restaurants joined delivery platforms \cite{McKinsey}. However, competition remains fierce among on-demand meal delivery (ODMD) platforms such as DoorDash, Uber Eats, JustEat, and Grubhub, making it crucial for these platforms to deliver exceptional customer experiences. Speed and reliability, particularly in meeting promised delivery times, are critical factors for maintaining customer satisfaction and loyalty \cite{liu2018online, koay2022model}. 

The operational flow of ODMD platforms involves multiple interdependent steps. 
In practice, once an order is placed on the platform, its details are sent to the restaurant for preparation. Meanwhile, a nearby courier receives the task and travels to the restaurant for pick-up. Once the order is ready, the assigned courier will have it delivered to the customer. 
Key metrics for success in this process are speed, availability, and punctuality. 
The customer desires less waiting time between the order placement and food arrival. The restaurant expects the meal to remain fresh when it arrives \cite{dai2020information,ulmer2021restaurant}. 
To maximize order fulfillment and maintain service quality, platforms must operate with high time efficiency, ensuring optimal fleet management to minimize unnecessary costs or resource wastage.
However, the operation of ODMD services is inherently complex due to highly dynamic and uncertain demand patterns.
Orders arrive stochastically on the platform, each with unique specifications. The arriving orders can be linked to any restaurant within the service network. Furthermore, customers have varying preferences for delivery times. Some customers prefer to receive their order as soon as possible, while others may specify a desired delivery time in the future.

Rather than relying solely on reactive, myopic adjustments based on current demand, ODMD platforms benefit significantly from short-term demand predictions to inform proactive real-time operations. Accurate demand forecasting helps platforms uphold user experience and optimize fleet utilization over time.
Recent research has explored advanced machine learning and deep learning approaches for spatial-temporal demand forecasting \cite{crivellari2022multi, yu2023short} and probabilistic predictions \cite{liang2023poisson}. Ensemble-learning models, including random forest and support vector machine, are also evaluated and compared with traditional time series approaches via a case study by Hess \textcolor{black}{et al.} \cite{hess2021real}. 

On-demand mobility and micro-delivery services in large cities face the significant challenge of managing numerous service locations with limited resources. 
\textcolor{black}{Therefore, in addition to accurate demand forecasting, clustering techniques are essential for managing the complexity of operations in large urban areas.}
By identifying emerging demand hotspots across neighborhoods, clustering provides actionable insights that help operators prioritize and relocate their limited fleet resources effectively  \cite{caggiani2017dynamic, chen2016dynamic}.
In addition, clustering simplifies operational optimization by grouping service zones with similar demand patterns, which can be used as management units for strategic planning.
This approach has been widely employed in on-demand services to facilitate various tasks, including facility location planning\cite{liu2022iterative}, route optimization \cite{prajapati2023clustering}, and fleet rebalancing \cite{lv2020hybrid, caggiani2017dynamic}.

\textcolor{black}{Practical challenges remain for real-time implementation of short-term demand forecasting and clustered insight generation.}
Meal delivery demand is highly complex, influenced by factors such as seasonality, weather, holidays, and special events, with spatial demand distributions fluctuating throughout the day and week.
Short-term demand forecasting must be computationally efficient and accurate to support frequent prediction updates for real-time operations. Static clusters based on historical data fail to meet the needs of dynamic operations. To be effective, clustering must account for both demand and spatial similarity, while also satisfying specific operational requirements. For instance, cluster generation for order bundling operations requires geographical contiguity and size constraints to maintain feasibility and optimize efficiency.

Addressing these challenges, this study introduces a novel predict-then-cluster framework designed to dynamically generate clustered insights from short-term demand predictions. The framework comprises two components: a demand forecasting step and a clustering step. The forecasting step leverages computationally efficient ensemble-learning models with multivariate features, including lagged demand, to capture both short- and long-term demand fluctuations. 
Guided by different spatial and operative requirements for the resulting clusters, we introduce Constrained K-Means Clustering (CKMC) and Contiguity Constrained Hierarchical Clustering with Iterative Constraint Enforcement(CCHC\textcolor{black}{-ICE}) that consider service zones' predicted demand and locations simultaneously.
This study evaluates the proposed framework through experiments in two case studies, using historical order data from the local delivery platform of a European and a Taiwanese city.
These experiments compare the performance of our ensemble-learning models with traditional time series prediction methods. Our findings demonstrate that the proposed models offer superior accuracy and computational efficiency. 
We analyze the clustering performance by comparing the clusters generated using predicted demand with those generated using actual demand. Our results highlight the benefits of incorporating distributional predictions to address demand uncertainties in clustering.
\textcolor{black}{To illustrate the practical implementation of the proposed framework, we conduct a simulation study on idle fleet rebalancing using demand predictions. The results show significant improvements in operational efficiency when predictions are integrated. This study highlights how short-term demand forecasts can guide proactive strategies, optimizing the efficiency of meal delivery services. }
\textcolor{black}{The coding implementation of the proposed predict-then-cluster framework, along with the datasets and experiments for the European and Taiwanese case studies, is shared with this publication.
To encourage applications and enhance reproducibility of this study, a comprehensive documentation is provided with the shared documents to clarity the usage of the framework.}

\textcolor{black}{The key contributions of our study are summarized as follows:}
\begin{itemize}
    \item \textcolor{black}{First, we introduce a short-term predict-then-cluster framework that sequentially combines forecasted demand information with clustering. 
    This novel framework is designed to dynamically produce clustered insights with the predicted demand information from the forecasting phase, which guides proactive operations of meal delivery platforms by identifying the future demand hotspots in the city.} \textcolor{black}{The frequently updated clusters together provide an overview of the evolving short-term demand dynamics within the meal delivery service network.}

    \item 
    Second, we propose enhancing short-term forecasting performance by incorporating lagged-dependent features, which capture sequential variations in demand to improve future predictions. These features enable high-quality short-term demand forecasts, even in scenarios with limited contextual information, such as the Taiwanese case.

    \item 
    Third, we address demand uncertainties through distributional forecasting using Quantile Regression Forest (QRF) and its lagged-dependent extension. These models exhibit robust forecasting performance across both case studies and enhance the predict-then-cluster framework by improving the quality of clustering outcomes through the integration of distributional demand predictions.
    
    \item 
    \textcolor{black}{
    To maximize resource utilization while improving the service level, short-term decisions play a crucial role in managing fleets more efficiently and effectively. To support these decisions, we introduce the Contiguity Constrained Hierarchical Clustering with Iterative Constraint Enforcement (CCHC-ICE), designed to enhance the efficiency of short-term operational decisions in meal delivery.
    This adaptive approach dynamically generates operationally viable clusters based on demand predictions, while ensuring compliance with contiguity constraints and user-defined requirements. 
    This approach combines a hierarchical clustering algorithm with an iterative constraint enforcement mechanism, ensuring that clusters meet both future demand similarity and operational constraints specified by operators. CCHC-ICE offers flexibility in incorporating various user-defined constraints to create tailored clustering solutions for diverse operational scenarios. In this study, we demonstrate its application by incorporating constraints such as maximum cluster size, minimum number of clusters, and dissimilarity thresholds.
    }
\end{itemize}

The remainder of this paper is structured as follows: Section \ref{sec:literature_review} reviews related literature on demand forecasting and clustering for on-demand services. Section \ref{sec:problem_description} formulates the predict-then-cluster problem and introduces the case studies. Section \ref{sec:method} describes the forecasting models and clustering algorithms. Experimental results are presented in Section \ref{sec:result}, followed by managerial insights in Section \ref{sec:disc}. Concluding remarks are provided in Section \ref{sec:conclusion}.

\section{Literature Review}
\label{sec:literature_review}
Demand forecasting has been an important area of interest in both fields of forecasting and operation research for various applications \cite{song2019review,suganthi2012energy,ghalehkhondabi2019review,fildes2019retail}. In subsection \ref{subset:challenges}, we highlight the benefits of high-resolution demand forecasting for the operations of on-demand delivery platforms and the challenges associated with high spatial and temporal forecasting granularity. Furthermore, we motivate how clustering can serve as an intermediary step to connect demand forecasts with real-time operations.
In subsection \ref{sec:literature_demand_forecast}, we discuss key studies on short-term demand forecasting, focusing on methods for handling complex seasonality and their applications, with a particular emphasis on recent advancements in forecasting for on-demand meal delivery (ODMD) services. Next, in subsection \ref{subset:LR_predict-cluster}, we review previous literature that presents a spatio-temporal predictive framework for on-demand services by integrating demand forecasting and clustering. 
Lastly, in subsection \ref{subsect:research_gaps}, we address the research gaps in the field and discuss how our study contributes to closing these gaps.

\subsection{\textcolor{black}{Challenges in operating on-demand meal delivery services with demand predictions}}
\label{subset:challenges}

Real-time operations, such as fleet rebalancing and order-matching, are central to the efficiency of on-demand meal delivery services. These operations are highly time-sensitive due to the limited lead time inherent in last-mile delivery. If the prediction horizon exceeds the operational completion time and is updated infrequently, the resulting demand forecasts may become outdated and unreliable, compromising optimization outcomes. 
Additionally, real-time rebalancing and order matching strategies that rely on specific locations may not fully leverage forecasted demand information for a larger service area. This indicates that demand forecasts should process a comparable temporal and spatial granularity to the target operations.
Effective and forward-looking decision-making for these real-time operations requires location-specific, up-to-date demand predictions alongside accurate fleet information.
Compared to forecasts over broader areas or longer horizons, fine-grained temporal and spatial demand forecasting provides more detailed and timely insights, supporting dynamic optimization in real-time operations. However, achieving this granularity introduces computational challenges, as forecasts must be generated quickly to remain actionable.

The design of a high-resolution, fast-computing short-term demand forecasting algorithm for a meal delivery platform should address several challenges, as outlined below.
First, demand patterns are often highly stochastic over short intervals in limited-sized service areas, resulting in noisy and intermittent data. Demand time series may exhibit high data sparsity, making it particularly susceptible to randomness, especially when the expected demand volumes are relatively low.
Second, demand often exhibits complex seasonality, with patterns influenced by seasonal factors such as time of year, day of the week, and hour of the day. These patterns vary significantly across different areas within a city \cite{hess2021real,yu2023short,liang2023poisson,crivellari2022multi}. For instance, business districts may see peak lunch-hour demand for sandwiches and coffee on weekdays, while residential areas may experience increased demand for pizza and Chinese food in the evenings or on weekends.
Furthermore, forecasting models must also be resilient to demand fluctuations caused by unforeseen events \cite{hess2021real, liang2023poisson}. Unforeseen circumstances in real life, such as sudden defect events and promotion events in a specific area\cite{liang2023poisson}. Beyond resilience, these models must be computationally efficient to support real-time predictions and adaptable to the evolving operational landscape of meal delivery platforms.
As platforms expand to new service areas and adjust their networks over time, forecasting methods must be generalizable, easy to update, and capable of performing well with limited observations \cite{hess2021real}. 
Maintaining zone-specific forecasting models offers additional flexibility, enabling independent updates without retraining the entire system or affecting predictions for other zones.

In addition to the forecasting challenges, operational optimization for numerous small service zones is also computationally challenging for platforms. 
In large cities like Shanghai or New York, platforms often manage hundreds of geographical units within their service networks. As managing these units individually is inefficient, it is a common strategy to group similar areas into clusters to streamline the planning process \cite{prajapati2023clustering}. Clustering similar zones reduces computational demands, allowing platforms to optimize operations sequentially at the cluster level rather than for individual zones. 
\textcolor{black}{Nonetheless, these clusters should be adaptively updated to reflect the evolving demand dynamics within the city. Changes in the demand landscape may arise from contextual factors such as weather variations or holidays.}

These reviewed challenges are not unique to meal delivery services but are shared by other on-demand urban mobility and logistics services. As such, investigating a generic predict-then-cluster framework offers broader applicability to similar challenges across various on-demand service domains.

\subsection{Methods and applications of short-term demand forecasting for on-demand services} 
\label{sec:literature_demand_forecast}
Dynamic demand predictions usually involve time-series data, comprising a sequence of data points measured at regular intervals over a period. Unlike cross-sectional data, time series observations typically vary with recent observations in sequence. 
Depending on the product, demand data may exhibit trends and seasonality patterns. 
Recent time-series applications on supply chain demand forecasting are reviewed by Seyedan et al. \cite{seyedan2020predictive}. 
Generally, there are two types of forecasting models based on predictive features, the univariate model with only the historical demand as input, and the multivariate model where external information (e.g., weather) is jointly used for prediction \cite{beeram2021time}. 

Traditional univariate time series models, such as exponential smoothing and its variants like Holt-Winter and SARIMA, have been successfully applied to short-term forecasting tasks, particularly when dealing with data exhibiting single seasonality. SARIMAX, an extension of SARIMA that incorporates external features, has further improved forecasting performance in applications like short-term load demand prediction \cite{cui2015short}. 
However, traditional models face significant challenges with data that exhibit complex or lengthy seasonal patterns. These limitations stem from the increasing parameter space required for longer seasonal periods, leading to higher computational costs, and the subjective nature of parameter selection. To address these challenges, De Livera et al. introduce a trigonometric exponential smoothing state-space model named TBATS \cite{de2011forecasting}. TBATS is specifically designed to address multiple seasonality while automatically optimizing model parameters. It has been effectively applied to forecasting tasks such as daily gas consumption \cite{naim2018effective} and demand at electric vehicle charging stations \cite{kim2021forecasting}.

With advancements in machine learning, non-linear multivariate methods such as random regression forest (RF), gradient boosting machines (GBM), and neural networks (NNs) have gained traction in short-term demand forecasting due to their ability to handle non-linear relationships and multivariate inputs \cite{aamer2020data}. 
Studies demonstrate their superior performance in tasks with complex seasonality. For instance, Dudek \cite{dudek2015short} applied random forests to forecast short-term load demand with multiple seasonal cycles. Albrecht et al. \cite{albrecht2021call} compared RF, SVM, KNN, and GBM in real-time call-center arrival forecasting, finding ensemble models consistently outperformed traditional time-series methods. These findings underscore the robust accuracy of ensemble learning methods, particularly when evaluated using metrics like mean squared error (MSE) and root mean squared error (RMSE).

Valuable insights can be drawn from short-term forecasting studies conducted for other on-demand services, such as ride-hailing and shared micro-mobility \cite{saadi2017investigation, qian2020short}.
Saadi et al. \cite{saadi2017investigation} investigate the performance of different machine learning models, including SVM, RF, GBM, and ANN-based regressions, using features related to pricing, weather, and traffic. Their findings highlight that ensemble-learning models, particularly GBM, achieve the highest prediction accuracy.
Qian et al. \cite{qian2020short} focus on 15-minute passenger demand forecasting for New York taxi services and propose a boosting Gaussian conditional random field model capable of generating robust point estimates and probabilistic predictions. Similarly, Noland \cite{noland2021scootin} applies RF to predict daily shared e-scooter and bike usage, considering the effects of weather, special events, and holidays.

In the domain of on-demand meal delivery services, research has primarily focused on fulfillment cycle time forecasting \cite{zhu2020order} and arrival time estimation \cite{hildebrandt2021supervised}, but there is growing interest in short-term demand forecasting. Hess et al. \cite{hess2021real} analyze hourly demand forecasting using data from an urban delivery platform in France. Their study reveals that exponential smoothing models outperform in scenarios with rich training data, while machine learning models excel with limited training data.  Crivellari et al. \cite{crivellari2022multi} introduce the Multi-target CNN-LSTM, a deep learning model for simultaneous next-hour demand forecasting across multiple service areas, updated every 15 minutes in their experiments. Building on this work, Yu et al. \cite{yu2023short} propose an attention-based convolutional LSTM model to capture the inter-location correlations and spatial variation among various meal delivery service areas in the city. Liang et al. \cite{liang2023poisson}  further extend the field by proposing a multi-task learning framework for distributional demand predictions. Their approach assumes a Poisson distribution for the order arrival process, estimating the associated parameters to quantify demand uncertainty.

\subsection{\textcolor{black}{Combined prediction and clustering framework for on-demand services}}
\label{subset:LR_predict-cluster}
\textcolor{black}{Integrating clustering with demand time series and geographical information across multiple service areas has received significant attention in recent literature. 
Many studies explore the advantages of cluster-then-predict framework to improve demand forecasting performance for on-demand mobility services\cite{chen2016dynamic, davis2016multi, feng2018hierarchical, kim2021spatial}. This hierarchical approach reduces the randomness in demand observations by aggregating demand from similar areas at a cluster level. Consequently, the demand predicted by a cluster-based forecasting model tends to be more robust than modeling individual areas with sparse demand. Chen et al. \cite{chen2016dynamic} propose a dynamic cluster-based forecasting approach to predict the over-demand probability for clustered areas in the city. They introduce a clustering method that considers geographical proximity, historical demand patterns of service areas, as well as the ongoing special events at the predicted time step. 
Kim \cite{kim2021spatial} introduces a spatial continuity-constrained hierarchical clustering method to generate clusters for bike-sharing traffic prediction. This study defines two grids to be spatially contiguous if the shortest path between them does not pass through any other grids with bike-sharing stations.}

\textcolor{black}{
On the other hand, the outcomes from clustering can be used to generate insights into the demand landscape within the service network over time \cite{liu2019exploring, caggiani2017dynamic}. Liu et al. \cite{liu2019exploring} use K-Means clustering with historical demand patterns and location data to identify which districts were historically more likely to have supply surplus and deficits at different periods for a ride-hailing service. To minimize the duration of time when vehicles are unavailable, Caggiani et al. \cite{caggiani2017dynamic} propose a dynamic clustering method to assist relocations of vehicles. Their approach jointly considers the geographical proximity, availability of vehicles, and received demand of service areas as clustering inputs. They show that the efficiency of fleet relocation is enhanced by utilizing dynamic zoning instead of static zoning.}

\subsection{Research Gaps}
\label{subsect:research_gaps}
\textcolor{black}{Upon reviewing the existing literature, we have identified several research gaps in demand forecasting for on-demand meal delivery.}
Firstly, the current literature on short-term demand forecasting is limited, \textcolor{black}{particularly for prediction horizons and update frequencies of less than an hour}. To the best of our knowledge, most studies focus on the hourly demand forecasting problem for ODMD \cite{yu2023short, crivellari2022multi, liang2023poisson, hess2021real}. \textcolor{black}{Demand forecasting for intervals shorter than an hour remains largely unexplored.}
Secondly, few efforts have been made to estimate the uncertainty of demand. The quantification of demand uncertainty has been proven to be the key to the success of operation policies by recent literature on on-demand mobility services \cite{lei2020efficient, huang2022one, guo2023data}. 
Liang et al. \cite{liang2023poisson} introduce a probabilistic prediction approach based on the Poisson assumption for the arrival of orders. However, this assumption might not fully capture real-life demand services, especially when demand changes during a period due to exogenous factors not considered in the model. As suggested by Lei et al. \cite{lei2014distribution}, it could be beneficial to explore non-parametric prediction models to address this limitation. 
Lastly, despite the increasing attention to hierarchical cluster-then-predict frameworks for demand forecasting, few studies have investigated predict-then-cluster approaches for dynamic cluster generation. 
Real-time operations based on dynamic zoning have shown to be more efficient \cite{caggiani2017dynamic}. However, operations like fleet rebalancing require lead time to take effect.
Thus, using demand predictions to generate forward-looking dynamic zoning further improves the operational efficiency of the system in the long run.
Additionally, geographical contiguity-constrained clustering methods remain underexplored in the context of on-demand meal delivery services. 
\textcolor{black}{Real-time operational efficiency can be improved by identifying dynamic clusters that satisfy both contiguity and user-specified constraints based on predicted short-term demand distributions across the city. This gap highlights the need for methodologies that integrate predictive demand models with flexible, geographically coherent clustering frameworks to support the unique challenges of on-demand delivery services.}

\section{Problem Description}
\label{sec:problem_description}
\textcolor{black}{In this section, we outline the research questions explored in this paper and introduce the datasets used for our empirical case studies. 
In Section \ref{subset:formulation}, we provide the formulation of our research problems, followed by the definitions of key terminologies in Section \ref{subset:definition}. Finally, Section \ref{subset:cases} provides an overview of our empirical case studies: a European case study and a Taiwanese case study.}

\subsection{\textcolor{black}{Research problem formulation}}
\label{subset:formulation}

\begin{figure}[h!]
    \centering
    \includegraphics[width=\linewidth]{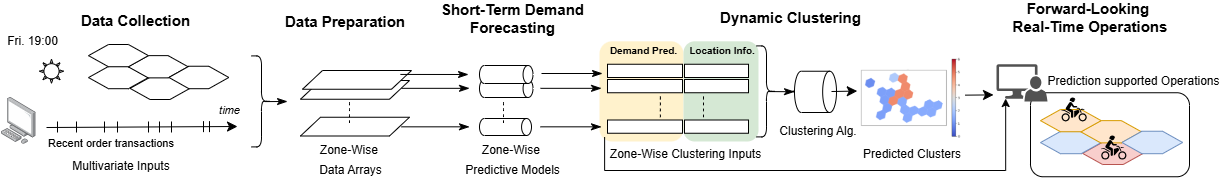}
    \caption{\textcolor{black}{Workflow of the predict-then-cluster framework for supporting real-time operations in meal delivery services.}}
    \label{fig:workflow}
\end{figure}

In this study, we aim to introduce a predict-then-cluster framework to accurately forecast demand across multiple service zones in the city and subsequently cluster similar zones to identify future demand coldspots and hotspots within the service network of a meal delivery platform. 
\textcolor{black}{Figure \ref{fig:workflow} provides an overview of the workflow for the proposed framework. The framework utilizes real-time multivariate inputs, such as date-time, weather, and order information, to identify recent demand patterns. Data from multiple sources is processed into zone-wise data arrays, which are fed into pretrained zone-specific predictive models for short-term demand forecasting.
The predicted short-term demand for each zone, combined with its location information, serves as input for dynamic clustering. The resulting clusters highlight areas with similar future demand, offering actionable insights to support real-time operational decisions, such as reallocating delivery couriers to zones with high anticipated demand.}

To summarize, the proposed framework comprises two components: short-term demand forecasting and dynamic clustering. The forecasting step aims to efficiently generate accurate short-term demand predictions using the latest demand and contextual information, leveraging a computationally efficient predictor to make optimal use of limited data. These demand predictions then serve as inputs for the clustering step, where service zones with similar predicted demands are grouped based on operational requirements such as geographical contiguity or proximity. The resulting clusters are designed to enhance delivery efficiency by meeting these practical constraints. Additionally, the dynamic clusters generated from predicted demand should closely resemble those formed using actual demand, thereby validating the robustness and accuracy of the predictive model.

\subsection{\textcolor{black}{Definitions}}
\label{subset:definition}

\textcolor{black}{We consider a platform's service region to be divided into local service zones.} In practice, these service zones are typically represented as equal-shaped grids within a city. Based on functionality, a service zone $z_i$ is called a \textbf{\textit{pick-up zone}} if \textcolor{black}{users can place orders from any restaurant within this zone, meaning couriers may pick up orders from $z_i$. On the other hand}, \textcolor{black}{a service zone} is a \textbf{\textit{destination zone}} if households in this zone can be served by the platform, i.e., this service zone can be listed as the destination of a meal delivery task. Forecasting tasks mainly concern the pick-up zones in this study. Hence, we define a set of pick-up zones $Z = \{z_1, z_2, \cdots, z_N\}$, where $N$ is the total number of pick-up zones.

\textcolor{black}{As discussed in Section \ref{subset:challenges}, to facilitate real-time operations, the meal delivery platform should generate frequently updated short-term demand forecasts.}
In this study, we use a 15-minute prediction horizon for short-term demand forecasting, with models updating demand predictions every 15 minutes.
Before model fitting, we pre-process historical order data to create a demand series for each pick-up zone. Assuming the platform updates demand predictions at 0, 15, 30, and 45 minutes past each hour during business hours, we reconstruct the order data into demand time series $[y^i_t, y^i_{t-1}, \cdots, y^i_1]$ by aggregating the number of orders related to each pick-up zone $i$ over the previous 15-minute intervals.
Similarly, the target variable for demand prediction $y^i_{t+1}$ is generated by counting the total number of orders received in the next 15-minute interval $t+1$ for zone $i$.

At the short-term demand forecasting step, predictions $\hat{y}^i_{t+1}$ are made for each pick-up zone $i$ in the next 15-minute interval $t+1$. \textcolor{black}{Our proposed demand forecasting models are trained specifically for each pick-up zone, using both global features and zone-specific features as inputs. Global features, denoted as $X_t$, have the same value for every pick-up zone at the current time interval $t$. In contrast, zone-specific features, denoted as $X^i_t$, have unique values for pick-up zone $i$. Therefore, the prediction of pick-up zone $i$ for the next time interval $t+1$ is generated by this zone's demand predictor using these features $\hat{y^i_{t+1}} = f^i(X_t, X^i_t)$. Depending on the selected algorithm, the demand forecasting model generates either point predictions or distributional predictions. While most algorithms output point predictions, distributional predictions are useful for estimating the uncertainty of the predicted demand.} 

\textcolor{black}{Demand predictions $\hat{Y}_{t+1} = \{y^1_{t+1}, y^2_{t+1}, \cdots, y^N_{t+1} \}$ are gathered as input to the clustering step. Depending on the requirements over the geographical proximity within clusters which are imposed by the target operations defined by the platform, we identify three possible cases:
\begin{enumerate}
    \item \textbf{No Geographical Proximity Required:} In this scenario, clustering can be performed solely based on the predicted demand of pick-up zones. A straightforward method is to classify zones using a simple thresholding approach to their predicted demand.
    \item \textbf{Geographical Proximity Required:} Many real-time operations of on-demand services can benefit from regional insights (as discussed in Section \ref{subset:challenges}). To support these operations, clusters are formed by jointly considering predicted demand similarity and geographical proximity of pick-up zones.
    \item \textbf{Geographical Contiguity Required:} In scenarios where geographical contiguity is necessary, such as for supporting order bundling operations, geographical contiguity is added as an additional constraint in the clustering process. This ensures that zones within the same cluster are contiguous.
\end{enumerate}
The proposed clustering methods target the last two scenarios requiring either geographical proximity or contiguity. At time interval $t$, the list of $K_t$ clusters $\Omega_t = \{\omega_1, \cdots, \omega_{K_t}\}$, where each cluster $\omega_k$ represents a set of pick-up zones resulting in it.   
}

\subsection{Empirical case studies}
\label{subset:cases}
To assess the performance of our proposed demand forecasting and clustering methods, we analyze the model's performance using meal delivery demand data from \textcolor{black}{two empirical case studies}. Before modeling, Exploratory Data Analysis (EDA) is performed to uncover the underlying patterns and reveal the potential data characteristics. In the following, we introduce the background of these case studies and highlight key insights from exploratory data analysis. 

\subsubsection{European Use Case}

The first case study features the meal delivery demand data from our industry partner, a leading meal delivery platform in Europe. This data contains simulated order transactions derived from real-life transactions in a European metropolitan city on our partner's platform. 
The dataset consists of a total of 188,162 historical orders spanning from April $1^{st}$, 2020 to September $14^{th}$, 2020. These orders are associated with 20 pick-up zones and 50 destination zones within the city. 
Each order transaction includes the date and time of order placement, as well as the \textcolor{black}{zonal addresses for} pick-up and delivery. The platform's regular service hours start at 10:30 AM and end at 9:30 PM every day. Orders received during business hours can thus be aggregated into 44 consecutive 15-minute intervals based on their placement time. Orders can only be \textcolor{black}{accepted by the platform} if the meal is to be collected from a restaurant within one of the designated pick-up zones and delivered to a household within a destination zone under service. 
The provided address information is hashed by Uber’s H3 geospatial indexing system at a resolution level of 8, with an average area of 0.737 $km^{2}$ per grid \cite{uberh3}. Thus, instead of being described by a pair of latitude and longitude, each address is represented as a hexagonal zone on the map. 

Although \textcolor{black}{the original} demand data is restricted from sharing, \textcolor{black}{a synthetic dataset created through the anonymization process described in \ref{appendix:anonymization} is provided to enhance the reproducibility of our study.}
In addition to the preliminary data analysis presented in this subsection, detailed findings from data analysis are also provided in \ref{appendix:dataset} to offer further insights into this European (EU) use case.

\begin{figure}[h!]
    \centering
    \includegraphics[width=0.55\textwidth]{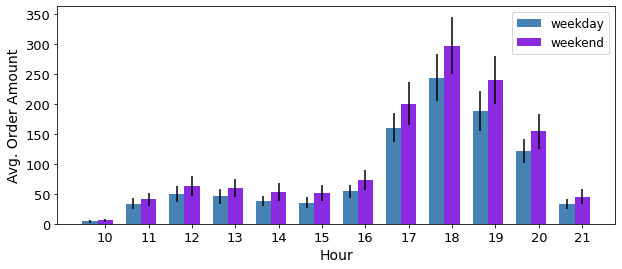}
    \caption{The average number of orders received in the \textcolor{black}{European} city during different hours \textcolor{black}{on weekdays versus weekends} , from April $1^{st}$, 2020 to September $14^{th}$, 2020.}
    \label{subfig:Daybars}
\end{figure}

\begin{figure}[h!]
    \centering
    \includegraphics[width=0.8\textwidth]{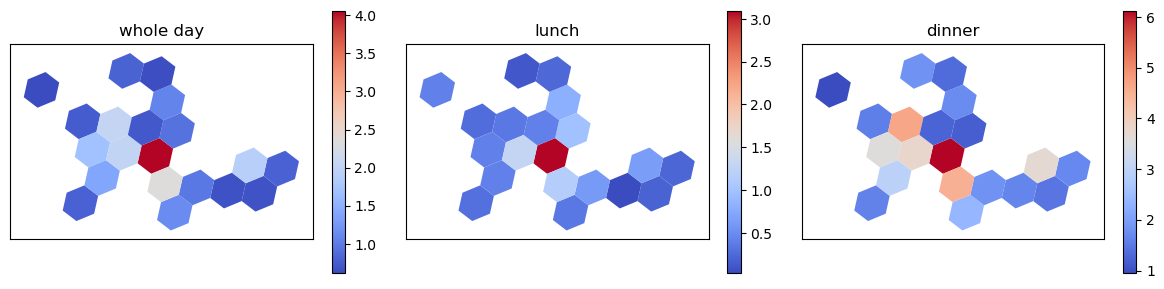}
    \caption{The heatmap visualization of the number of orders received by different pick-up zones \textcolor{black}{from the European use case} per 15-minute intervals, averaged over three time periods: whole day, during lunch times, and dinner times. \textcolor{black}{The heatmaps use a color scale to indicate the density of orders, with hotter colors representing higher zonal order volumes.}}
    \label{Fig:heatmap_15_min_orders}
\end{figure}

\textcolor{black}{The preliminary data analysis shows} an average daily order volume ranging between 800 and 1400. It also suggests a recurring weekly periodic pattern with relatively lower demand volumes from Monday to Thursday, and generally higher volumes from Friday to Sunday.
To further explore the weekly and daily seasonal patterns in this data, we compare the average total number of order arrivals per business hour received in the European city. We categorize the week into weekdays (Monday to Thursday) and weekends (Friday to Sunday). Friday is included in the weekend category as customers often treat Friday evening as the start of the weekend, with most orders arriving during dinner time. 
Figure \ref{subfig:Daybars} shows the average number of orders received in the European city at different hours, separated by weekdays and weekends. This bar plot indicates that the demand volume exhibits a dual-peak pattern around normal lunch (12 PM) and dinner (6 PM) times for both weekdays and weekdays. The dinner peak has a significantly higher number of orders compared to the lunch peak, while the lunch peak does not show an evidently higher demand than the afternoon period. 
Overall, demand is higher on weekends compared to weekdays, especially in the evening.

In addition to weekends, holidays may also significantly impact the demand for meal delivery services \cite{tong2020will,chen2020influence}. 
There are 8 national holidays between April $1^{st}$ and September $14^{th}$, 2020. Analyzing the number of orders received on these dates, we found that 5 out of the 8 holidays had more orders than 80\% of the corresponding weekdays. Notably, 4 out of these 5 high-demand holidays fell on weekdays. Our findings suggest that holidays have a positive influence on the demand for food delivery in the city. 

Next, we explore the demand aggregated at 15-minute intervals. In Figure \ref{Fig:heatmap_15_min_orders}, we present a heatmap visualization to explore the spatial demand patterns for the whole day, during lunch (11 AM-2 PM) and dinner (5 PM-9 PM) periods. The heatmaps reveal that a central pick-up zone consistently exhibits the highest average demand compared to other zones. Additionally, the spatial demand distribution varies between lunch and dinner periods, with certain pick-up zones experiencing a more significant increase in demand during dinner times compared to lunch times.
\textcolor{black}{To examine the sparsity of demand data in a demand series aggregated at 15-minute intervals, we analyze the sparsity ratio, which shows the proportion of observations with zero values. In the European use case, zero values account for an average of 47.0\% of total observations in the 15-minute aggregated demand series, with a maximum sparsity ratio of 67.2\% among the pick-up zones.}

\textcolor{black}{The weather conditions, such as precipitation of rain and snow, have been shown to significantly impact the demand for meal delivery services \cite{yao2023weather}.} \textcolor{black}{For the European use case, our analysis suggests that light precipitation may attract more orders compared to dry conditions, while heavy precipitation tends to result in lower demand volumes.}

\subsubsection{Taiwanese Use Case}

\textcolor{black}{In addition to the European use case, we conducted a second case study for the city of Taiwan, utilizing the open-sourced order transaction data by \textit{Delivery Hero}, a global meal delivery service provider \cite{assylbekov2023delivery}. Since 2012, Delivery Hero has been operating meal delivery services in the city of Taiwan. Delivery Hero has been operating meal delivery services in Taiwan since 2012. According to the dataset from Assylbekov et al., \cite{assylbekov2023delivery}, the platform facilitates more than 22,000 orders per day on average, covering a service area of 600 $\text{km}^2$ in Taiwan.}

\textcolor{black}{
The dataset has been anonymized in advance by the platform to ensure privacy. 
Each order transaction in the data includes the pick-up and drop-off locations, the placement time, and day-of-the-week information. Unlike the European use case, specific date information is unavailable in this Taiwanese dataset. 
The pick-up and drop-off locations are hashed into zonal addresses using a geohash procedure at precision level 5. There are 25 pick-up zones in this dataset, each covering a squared area of approximately 24 $\text{km}^2$. The historical orders spanned a three-month period, consisting of 1.64 million orders from the first 76 days (equivalent to ten weeks) used for training data, and 0.36 million orders from the last 14 days (a two-week period) for testing. }

\begin{figure}[h!]
    \centering
    \includegraphics[width=0.75\textwidth]{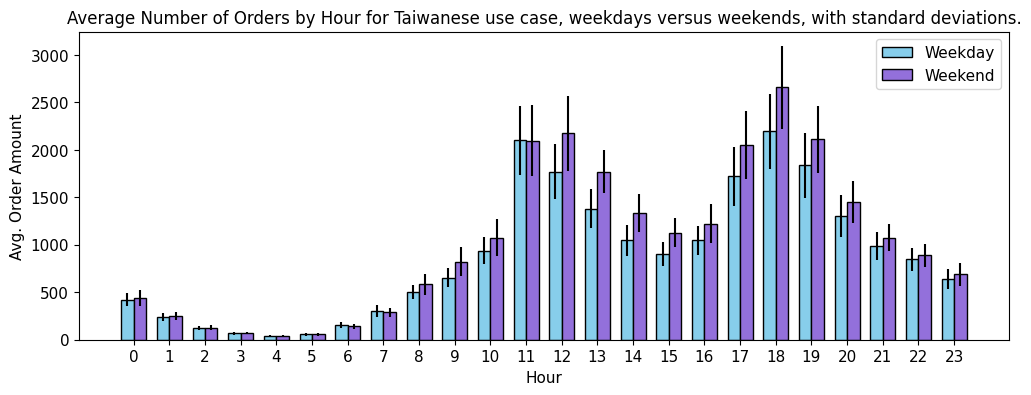}
    \caption{\textcolor{black}{The average number of orders received in Taiwan during different hours on weekdays versus weekends, was calculated using a dataset gathered over a three-month period. Error bars represent the standard deviation per hour across different days.}}
    \label{fig:Taiwan_demand_pattern}
\end{figure}

\begin{figure}[h!]
    \centering
    \includegraphics[width=0.90\textwidth]{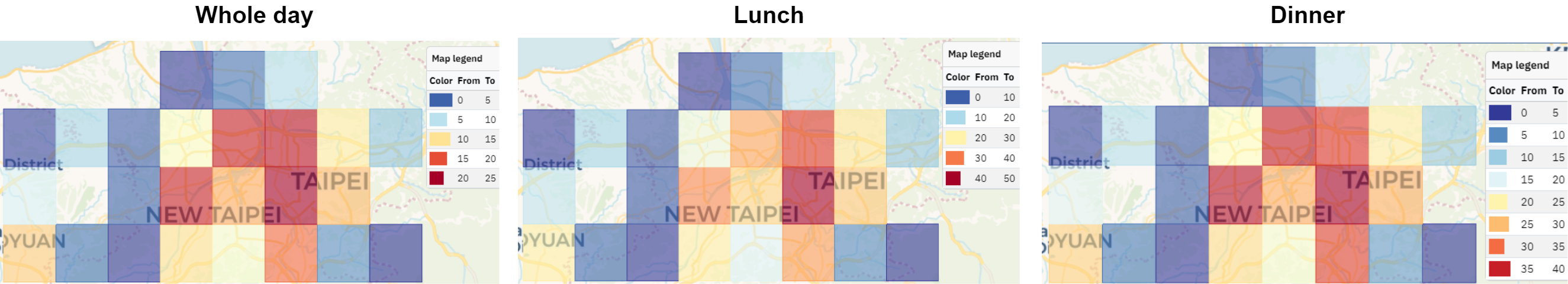}
    \caption{\textcolor{black}{The heatmap visualization of the number of orders received by different pick-up zones from the Taiwanese use case per 15-minute intervals, averaged over three time periods: whole day, during lunch times, and dinner times. The heatmaps use a color scale to indicate the density of orders, with hotter colors representing higher order volumes.}}
    \label{fig:heatmap_TW}
\end{figure}

\textcolor{black}{Splitting the order data into weekdays and weekends as before, Figure \ref{fig:Taiwan_demand_pattern} presents a bar plot of the average number of orders received by hour throughout a day. With the delivery service in Taiwan operating 24 hours a day, the order volume exhibits a dual-peak pattern, peaking at lunch and dinner hours. The average order volumes for lunch and dinner hours on weekdays and weekends are comparable, with a slightly higher volume around 6 PM on weekends compared to the other peak hours. Similar to the European case, the hourly order volumes on weekends are generally higher than those on weekdays.}

\textcolor{black}{To explore the spatial demand patterns in the Taiwanese use case, we visualize the average demand aggregated per 15-minute interval for each pick-up zone in Figure \ref{fig:heatmap_TW}. Similar to the analysis conducted for the European city, the heatmaps are generated for the whole day level, lunch times, and dinner times. The heatmaps reveal that the relative demand levels among zones remain largely consistent during lunch and dinner times. However, some zones at the center of the service network experience a more significant increase in demand during dinner hours compared to other pick-up zones. Additionally, the average number of orders received per 15 minutes is significantly higher compared to the European case, due to the larger service area covered per pick-up zone. Consequently, the sparsity ratios per 15-minute aggregated demand series are much lower in the Taiwanese use case, with only 18.3\% of total observations being zero in the pick-up zone on average.}

\section{The Short-Term Predict-then-Cluster Framework}
\label{sec:method}

This section outlines the design of the proposed short-term demand predict-then-cluster framework for on-demand meal delivery services. 
\textcolor{black}{Section \ref{subsect:forecasting_framework} introduces the models selected for short-term demand forecasting within this framework.} For the dynamic generation of clusters based on predicted demand, two clustering approaches, \textit{Constrained K-Means Clustering} (CKMC) and \textit{Contiguity Constrained Hierarchical Clustering \textcolor{black}{with Iterative Constrained Enforcement}} \textcolor{black}{(CCHC-ICE)}, are introduced in Section \ref{subsect:clustering}.

\subsection{\textcolor{black}{Short-term demand forecasting for on-demand meal delivery services}}
\label{subsect:forecasting_framework}

In this subsection, we introduce the selected tree-based ensemble learning models and benchmark regressors for deterministic and distributional demand forecasting of meal delivery services \textcolor{black}{in Sections \ref{subset:forecasting_models_point} and \ref{subset:forecasting_models_quantile}, respectively}. Next, we describe the proposed lagged-dependent extension for ensemble learning predictors in Section \ref{subset:LD_models}.

\subsubsection{Deterministic short-term demand forecasting models}
\label{subset:forecasting_models_point}
Tree-based ensemble-learning methods, such as random forest regression (RF) and XGBoost regression, have shown promising performance in deterministic hourly demand forecasting for meal delivery services\cite{hess2021real} and other short-term demand forecasting applications \cite{he2020day, albrecht2021call,saadi2017investigation}. 
These methods are recognized for their robustness, interpretability, and efficiency, outperforming neural networks in various forecasting tasks \cite{vairagade2019demand, lu2019performance}.
\textcolor{black}{Ensemble methods like RF and XGBoost handle challenging datasets, such as intermittent time series with frequent zeros, common in short-term meal delivery demand. By using decision trees as base learners, they partition data into scenario-specific subsets based on input features, effectively capturing complex patterns. This capability enables them to deliver robust and accurate predictions, even for irregular or extreme demand patterns where conventional regression methods often struggle.}
Therefore, this study focuses on RF and XGBoost for deterministic forecasting. Detailed formulations and pseudocode for RF and XGBoost are provided in \textcolor{black}{\ref{appendix:rf} and \ref{appendix:xgboost}, respectively}.

To benchmark the performance of the ensemble methods, we include widely used time series models such as SARIMA, SARIMAX, and TBATS, which have demonstrated state-of-the-art performance in demand forecasting tasks across various domains \cite{kaffash2021big, kumar2015short, shuai2022relationship, nosal2014effect, chiang2011forecasting}. \textcolor{black}{Detailed descriptions of these benchmarks are presented in \ref{appendix:point_benchmarks}.}
In practice, many meal delivery platforms do not yet implement demand forecasting. A simple yet naive alternative is to use the most recent observation as the prediction for the next time interval. This approach, referred to as the \textbf{myopic} predictor, is also included as a benchmark for deterministic demand forecasting. 

\subsubsection{Distributional short-term demand forecasting models}
\label{subset:forecasting_models_quantile}
Distributional demand predictions offer two key advantages. First, they allow the use of the median of the predicted distribution as a point estimate. Unlike traditional regression methods that default to predicting the mean, median-based predictions are less influenced by extreme values or outliers, resulting in more robust performance. Second, distributional predictions quantify demand uncertainty, providing critical insights that enable platforms to optimize operational policies more effectively.
To leverage these benefits, we apply quantile regression forest (QRF), a distributional forecasting variant of RF introduced by Meinshausen \cite{meinshausen2006quantile}. QRF generates robust point predictions while also capturing uncertainty through demand quantiles. It has demonstrated strong performance in various applications, including short-term electricity demand forecasting \cite{xing2020load} and online grocery retailing demand \cite{ulrich2021distributional}. \textcolor{black}{Detailed pseudocode and explanations for QRF are provided in \ref{appendix:qrf}.}

\textcolor{black}{As a benchmark, we include the \textbf{Seasonal Quantile (Seasonal)} predictor to evaluate the distributional forecasting performance of QRF methods. The Seasonal predictor segments historical demand data by hour and day of the week to create time-dependent empirical distributions for each pick-up zone. Quantile predictions are then generated based on these seasonally conditioned distributions, providing a baseline for evaluating the effectiveness of the proposed methods.}

\subsubsection{Lagged-dependent ensemble-learning \textcolor{black}{models for short-term demand forecasting}}
\label{subset:LD_models}

Although seasonal variables can be selected from past data observations, there may be uncovered factors that can significantly \textcolor{black}{impact} demand levels. For instance, demand could rise because of restaurant promotions or during the break of a popular football game. The demand may also decrease temporally due to platform or restaurant decisions to limit demand because of insufficient staff. To \textcolor{black}{address} this issue, we propose to include the lagged-dependent (LD) features, which are the recent demand observations, in the decision-tree-based ensemble-learning models.

The \textcolor{black}{inclusion} of past observations has shown potential in traditional time series and machine learning methods.
The autoregressive (AR) model identifies a linear relation between the target variable $y_t$ and its recent observations. The regression function of AR only includes the previous observations, a constant, and the error term.
Subsequence time series clustering aims to \textcolor{black}{detect} reoccurring patterns of time series subsequence and utilize these patterns for prediction. The approach uses a sliding window to divide a long time series into subsequences, which are then clustered based on similarity measurements \cite{aghabozorgi2015time}. Both models leverage groups of neighboring observations as proxies for predicting the target variable.

Inspired by AR and time series subsequence clustering models, we propose incorporating four lagged-dependent terms $y_{t}, y_{t-1}, y_{t-2}, y_{t-3}$ as additional temporal features to capture the demand fluctuations. 
Trained with these additional lagged-dependent features, extended ensemble-learning models LD-RF, LD-XGBoost, and LD-QRF, are introduced in our study.
The advantage of LD extension is that the model can leverage insights summarized \textcolor{black}{from} recent demand observations to enhance forecasting performance, without requiring extensive efforts in model training and data collection. 
Compared to subsequence time series clustering, LD extension does not need to train a separate clustering method for time series subsequences. Additionally, it offers a preferable solution over SARIMA models by \textcolor{black}{retaining} fast computing, non-linearity, and robustness properties of decision-tree-based ensemble-learning methods.

\subsection{Dynamic predict-then-cluster framework for demand cold- and hotspots forecasting}
\label{subsect:clustering}

In this study, we introduce a predict-then-cluster framework to generate dynamic clusters for identifying predicted demand cold- and hotspots within the service network of a meal delivery platform. \textcolor{black}{These outcomes offer a dynamic spatial overview of short-term demand across the city, adapting to fluctuations in demand.}
\textcolor{black}{Firstly,} short-term demand predictions for the next 15 minutes are generated according to one of the forecasting models discussed in the previous subsection \ref{subsect:forecasting_framework}. \textcolor{black}{Next,} the predicted demand and the geographical information for all zones are jointly taken as input to the clustering process. This dynamic predict-then-cluster approach groups similar pick-up zones together, providing predictive insights into demand clusters for the upcoming 15 minutes.

In this research, we introduce two types of clustering approaches to apply in the predict-then-clustering framework: Constrained K-Means Clustering (CKMC) and Contiguity Constrained Hierarchical Clustering with Iterative Constraint Enforcement (CCHC-ICE). These approaches are particularly useful for generating dynamic, map-based insights to support managerial decision-making.
CKMC groups similar zones into larger, cohesive units for managerial operations by enforcing a minimal cluster size constraint. This allows operators to dynamically define operational areas and compare predicted demand across the city. However, its limitation lies in potentially forming geographically disconnected clusters, which can hinder efficiency in scenarios requiring physical connectivity.
\textcolor{black}{
On the other hand, CCHC-ICE ensures geographical connectivity by enforcing contiguity and user-specified constraints throughout the clustering process. This method iteratively forms clusters that preserve zonal contiguity and predicted demand similarity, making them both operationally meaningful and geographically coherent. The resulting clusters can act as seamless geographical units for operations, such as meal delivery order bundling and routing.}

\subsubsection{Constrained K-Means Clustering}

By measuring the similarities among clusters with both location and predicted demand information as inputs, \textcolor{black}{CKMC helps monitor the spatial distribution of over-demand/under-supplied areas.}
\textcolor{black}{CKMC utilizes K-Means clustering to group zones into clusters.
This approach partitions a high-dimensional space spanned by data points into clusters by identifying representative centroids for each cluster. The algorithm assigns each data point to the nearest centroid, aiming to minimize the within-cluster variance.}
\textcolor{black}{K-Means assumes the clusters to be approximately spherical and evenly sized. It is suitable for scenarios where managerial clusters are created to direct the platform on resource allocation and improve operational efficiency, such as fleet rebalancing and demand management.}

\textcolor{black}{In CKMC, the geographical proximity between two pick-up zones is measured by the distance between their grid centers.} For each zone, the latitude and longitude of its grid center are included as \textcolor{black}{static} geographical features. \textcolor{black}{The predicted demand features for clustering are dynamic, which are provided from the previous prediction step within the predict-then-cluster framework for the target 15-minute time window $t$ for each pick-up zone. Depending on the forecasting approach in the prediction part of the predict-then-cluster framework, the type of predicted demand features may vary.} 
Particularly, demand uncertainty can be taken into account at the clustering phase by using multiple quantile predictions from QRF and LD-QRF.  
\textcolor{black}{The distance measure for CKMC is defined as,}
\begin{equation} \label{eq:clustering_distance}
    d(x_{i}, x_{j}) = \sqrt{\sum_{k=1}^{d} w_k (x_{i,k} - x_{j,k})^2},
\end{equation}
\textcolor{black}{given input feature vectors of any two pick-up zones $i$ and $j$. Predefined weight parameter $w_k$ is allocated to each feature $k$ to control its relative importance in the distance measure.}

\textcolor{black}{Distance measures are utilized in the clustering process to assess the similarity between elements. The smaller the distance between two elements, the more similar they are.
For the predict-then-cluster experiment with QRF and LD-QRF, we consider applying} 0.25, 0.50, and 0.75 quantile predictions of each pick-up zone to measure the predicted demand similarity. Equal weights are allocated to each geographical and predicted demand feature. \textcolor{black}{The distance measure can be further simplified as \[d(x_{i}, x_{j}) = \sqrt{(\textit{lat.}_{i} - \textit{lat.}_{j})^2 + (\textit{lng.}_{i} - \textit{lng.}_{j})^2 + \sum_{q=0.25, 0.50, 0.75}(\hat{y}^{q}_{i} - \hat{y}^{q}_{j})^2}.\]} 
\textcolor{black}{Operator users of the predict-then-cluster framework can also finetune the hyperparameters, including the number of quantile predictions and the corresponding weights allocated to} these features, based on the optimization performance in specific use cases. For the other experiment with point predictions as input to clustering, the weight for the predicted demand feature is tripled to keep the relative importance of location input the same as before. \textcolor{black}{In this case, the distance measure follows, \[ d(x_{i}, x_{j}) = \sqrt{(\textit{lat.}_{i} - \textit{lat.}_{j})^2 + (\textit{lng.}_{i} - \textit{lng.}_{j})^2 + 3 \times (\hat{y}_{i} - \hat{y}_{j})^2}. \]}
Before clustering, all features are normalized to the same scale.
Silhouette coefficient is a clustering performance metric, calculated based on the distance measure \textcolor{black}{$d(x_{i}, x_{j})$}.
It is applied to compare the similarity of a sample within its current cluster to other clusters \cite{rousseeuw1987silhouettes}.

\subsubsection{Contiguity Constrained Hierarchical Clustering \textcolor{black}{with Iterative Constraint Enforcement}}
\label{sect:CCHC-ICE}
\textcolor{black}{Along with cluster-specific conditions such as the number of clusters and cluster sizes, }
geographical contiguity is often a critical requirement for supporting operations.
This condition ensures connectivity within each cluster, guaranteeing that all zones from the same cluster are directly accessible to one another. 
However, while methods like CKMC consider geographical proximity, they do not guarantee such contiguity. As a result, zones with similar demand levels may be grouped together even if natural barriers, such as road networks, waterways, or mountainous regions, render them inaccessible from one another. This lack of contiguity can hinder applications requiring coherent and seamless cluster configurations, such as order batching or route planning, where continuous and logical path connectivity is essential.
To address these limitations, we introduce \textcolor{black}{an adaptive Contiguity Constrained Hierarchical Clustering with Iterative Constraint Enforcement (CCHC-ICE) framework designed for dynamic predicted demand cluster generation under user-defined requirements. Accepting short-term predictions as input, this approach integrates an iterative constraint enforcement mechanism with a contiguity constrained hierarchical clustering algorithm, ensuring that resulting clusters satisfy both future demand similarity and user-specified criteria regarding geographical contiguity and other properties.}

\subsubsection*{\textcolor{black}{\textbf{Contiguity Constrained Hierarchical Clustering}}}
Contiguity Constrained Hierarchical Clustering (CCHC) extends traditional hierarchical clustering by incorporating contiguity constraints, ensuring that clusters meet specific requirements for geographical connectivity \cite{guenard2022hierarchical}.
The process of agglomerative hierarchical clustering can be illustrated by a dendrogram. It begins with each data point as an individual cluster. At each iteration, the pair of clusters with the highest similarity (or shortest distance) are merged. This process continues until only one cluster remains or a stopping criterion is met \cite{murtagh2012algorithms}. In CCHC, additional constraints ensure that merging two clusters does not violate contiguity requirements. For example, zones within a cluster must remain directly reachable without leaving the cluster \cite{guenard2022hierarchical}. 

\textcolor{black}{The incorporation of contiguity constraints within hierarchical clustering has been explored in the literature. Gu\'{e}nard and Legendre \cite{guenard2022hierarchical} introduce a general framework for CCHC with a Lance-Williams algorithm, where an adjacency matrix captures contiguity relationships and is updated iteratively via a dissimilarity matrix.
Chavent et al. \cite{chavent2018clustgeo} introduce ClustGeo, which balances feature-based and geographical distances using two dissimilarity matrices, offering flexible clustering with soft spatial constraints. C{\^o}me \cite{come2024bayesian} presents a Bayesian clustering approach incorporating prior knowledge (e.g., expected clusters) to compute maximum a posteriori partitions, generating dendrograms that account for both spatial and probabilistic constraints.
Guo \cite{guo2008regionalization} introduces REDCAP (Regionalization with Dynamically Constrained Agglomerative Clustering and Partitioning), addressing limitations in earlier CCHC models that only consider static, first-order contiguity. REDCAP dynamically updates the contiguity matrix after each merge operation, incorporating full-order contiguity constraints. By considering the relationships among all elements in the merging clusters, REDCAP ensures spatial coherence throughout the iterative clustering process \cite{guo2008regionalization}.}

\subsubsection*{\textcolor{black}{\textbf{Iterative Constraint Enforcement}}}
\textcolor{black}{Existing models for Contiguity-Constrained Hierarchical Clustering (CCHC) primarily focus on enforcing geographical contiguity during the merging process, often overlooking other critical cluster characteristics. Most frameworks \cite{guenard2022hierarchical, chavent2018clustgeo} rely on static adjacency matrices, prioritize minimizing dissimilarity and maintaining spatial adjacency.  While REDCAP \cite{guo2008regionalization} supports dynamic contiguity updates, its scope is primarily focused on spatial applications, limiting its versatility for other operational or domain-specific requirements.}

\textcolor{black}{To address these limitations, we propose a CCHC framework with an Iterative Constraint Enforcement (ICE) mechanism embedded in the agglomerative hierarchical clustering process. This mechanism dynamically verifies and enforces not only contiguity but also user-defined constraints, such as size, shape, or attribute homogeneity, at each step of the clustering process. By integrating these considerations, the proposed framework ensures that the resulting clusters are operationally viable, geographically coherent, and aligned with application-specific requirements.}

In the \textcolor{black}{CCHC-ICE} framework, merging conditions between clusters are dynamically updated to account for both geographical contiguity and user-defined constraints, based on the current composition of each cluster. A pair of clusters is eligible for merging only if the resulting merged cluster satisfies all specified constraints. This ensures that, at each step, only pairs of clusters meeting the criteria are considered, maintaining adherence to constraints throughout the clustering process.
The framework enforces these constraints by iteratively updating a constrained distance matrix $D_z$, which measures the zone-wise dissimilarities under constraints. 
The calculation of $D_z$ involves updates to two symmetric binary matrices $C_p$ and $C_e$ at each iteration.  

Constraint condition matrix $C_p$ specifies whether two zones respectively belong to a pair of feasible contiguous clusters under all constraints.
Upon initialization of the algorithm, a predefined symmetric binary matrix is provided to represent the adjacency relationships among zones, where two zones are considered adjacent if they share a common border on the map. \textcolor{black}{Following the definition in Gu\'{e}nard and Legendre \cite{guenard2022hierarchical},} geographical contiguity between two clusters is established if at least one zone from one cluster is adjacent to a zone in the other cluster. Additionally, for two clusters to be considered qualified for merging, they should also meet the user-specific constraints. For instance, if maximal cluster size is defined, the combined size of two clusters shall be under this threshold.
The algorithm defines $C_p [i,j] = 1$, if zones $i$ and $j$ belong to a pair of clusters qualified for merging under constraints, otherwise $C_p [i,j] = 0$.

To accelerate clustering computations in the iterative framework, the matrix $C_e$ is updated at each step to identify whether a pair of zones currently belong to the same cluster. For a pair of zones $i$ and $j$ from the same cluster, $C_e [i,j] = 1$, otherwise $C_e [i,j] = 0$.
This effectively transfers the clustering outcomes from the previous iteration to the current one, allowing the algorithm to bypass calculations from previous agglomerative hierarchical clustering steps and thereby accelerating the process.

Given zones $i$ and $j$, a distance of zero is assigned if they already belong to the same cluster. Elsewise, the zone-wise distance $D_z [i,j]$ follows the regular distance measure $d(x_i,x_j)$ from Equation \eqref{eq:clustering_distance} with each weight parameter $w_k = 1$.
If their respective clusters shouldn't be merged in the next step according to the constraints, the zone-wise distance is set to $+ \infty$. The calculation of $D_z [i,j]$ is summarized below:
\begin{equation}
    D_z [i,j] = 
    \begin{cases}
        0 & \quad C_e [i,j] = 1 \; \text{and} \; C_p [i,j] = 0,\\
        d(x_i,x_j) & \quad C_e [i,j] = 0 \; \text{and} \; C_p [i,j] = 1,\\
        + \infty & \quad C_e [i,j] = 0 \; \text{and} \; C_p [i,j] = 0.
    \end{cases}
\end{equation}

At each iteration, $D_z$ is updated and input into a standard Agglomerative Hierarchical Clustering (AHC) algorithm \cite{murtagh2012algorithms}.
At each iteration, clusters with the smallest cluster-wise distance $\bar{d}$ are merged. The cluster-wise distance is defined as the average distance between all pairs of zones in the two clusters, consistent with standard AHC criteria \cite{murtagh2012algorithms}. 
Additionally, a violation detector is incorporated to verify the contiguity conditions of the resulting clusters at the end of each iteration. If a violation is detected, the clustering process terminates, and the last valid set of clusters is returned.

\subsubsection*{\textcolor{black}{\textbf{Example operative constraints integration}}}

\textcolor{black}{Tailoring to the case study conducted in this research, we showcase the integration of three types of user-specific constraints into the CCHC-ICE framework: a cluster dissimilarity threshold, a minimum number of clusters, and a maximum cluster size.}
To limit the merging of dissimilar clusters, a cluster dissimilarity threshold $D_{max}$ can be included. If the cluster-wise distance in the next merging step exceeds this threshold, the clustering process terminates, and the results from the previous iteration are returned. This ensures that clusters remain homogeneous in terms of their dissimilarity measures.
The minimum number of clusters, $K_{\text{min}}$, \textcolor{black}{is useful for operations when it is necessary to maintain a certain level of granularity in the clustering output, such as in applications requiring a minimum number of operational units.}
The iterative clustering process of \textcolor{black}{CCHC-ICE} terminates when the current number of clusters reaches $K_{\text{min}}$ or when a contiguity constraint violation is detected.
\textcolor{black}{Finally, limiting the maximum cluster size is essential where overly large clusters may hinder operation efficiency. For instance, in the case of meal delivery and other last-mile logistic services, clusters as operation units should be within manageable size }
Given a maximum cluster size of $s_{\text{max}}$, the merge between two clusters is only feasible if the sum of their cluster size does not exceed $s_{\text{max}}$. 
\textcolor{black}{Pseudocodes \ref{alg:CCHC} and \ref{alg:CCHC_helper} in \ref{appendix:cchc} provide a detailed outline of the implementation for the specified constraints within the CCHC-ICE framework.}

\begin{figure}[h!]
    \centering
    \includegraphics[width= 0.75\textwidth]{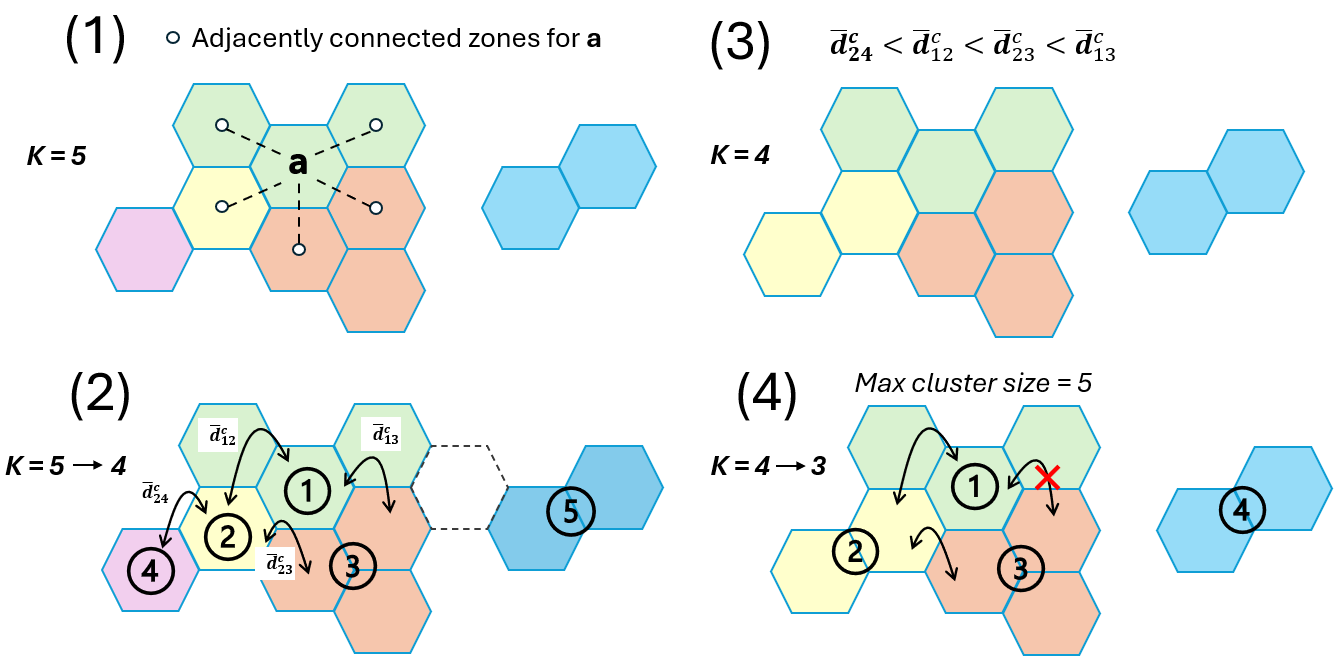}
    \caption{\textcolor{black}{Illustration of the iterative clustering process in the Contiguity Constrained Hierarchical Clustering \textcolor{black}{with Iterative Constraint Enforcement (CCHC-ICE)} algorithm: $K$ is the number of clusters at the current iteration. \textit{Max cluster size} refers to the maximum number of zones within a cluster specified in the CCHC\textcolor{black}{-ICE} algorithm.} $\bar{d}^c_{mn}$ is the average cluster-wise distance between clusters $m$ and $n$. Arrow $\rightarrow$ denotes the computation process for one-step \textcolor{black}{merging} within the agglomerative iterative clustering algorithm.}
    \label{fig:cchc}
\end{figure}

Figure \ref{fig:cchc} illustrates the iterative cluster merging process in the CCHC-ICE algorithm.
subplot (1) displays the adjacency relationships for zone $a$. 
Subplot (2) shows that cluster \circled{1} (green) is contiguous with clusters \circled{2} (yellow) and \circled{3} (orange), forming feasible pairs for merging based on contiguity constraints. 
Subplots (2) and (3) together depict a single iteration of agglomerative hierarchical clustering (AHC), where clusters are merged in order of increasing cluster-wise distances among feasible, contiguous pairs. This one-step merging happens for the pair exhibiting the smallest cluster-wise distance.
Subplot (4) shows an example an example where the merge between clusters 
\circled{1} (green) and \circled{3} (orange) is seen infeasible due to the maximum cluster size constraint being exceeded.

\section{Experiments and Results}
\label{sec:result}
This section presents \textcolor{black}{the design} and computational results of the short-term predict-then-cluster experiments for on-demand meal delivery case studies. 
Subsection \ref{subsec:experiment_design} details the experiment design, \textcolor{black}{including the model training and prediction generation process, input features, and evaluation scheme of the experiments.} Section \ref{subsec:numerical} compares and analyzes the computational efficiency of our selected forecasting and clustering algorithms. Subsection \ref{subsec:forecasting_result} presents the performance of deterministic \textcolor{black}{and distributional} demand forecasting for the \textcolor{black}{European and Taiwanese use cases. Lastly, Subsection \ref{subsec:clustering_result} presents the results of dynamic clustering using demand predictions.} 

\subsection{\textcolor{black}{Design of Case Studies}}
\label{subsec:experiment_design}
\textcolor{black}{This subsection begins by detailing the input features selected for demand forecasting. It then provides an overview of the model training and prediction generation processes, followed by an explanation of the evaluation schemes used for both demand forecasting and clustering experiments.} 

\subsubsection{Features for short-term demand forecasting}

\begin{figure}
    \centering
    \includegraphics[width= 0.9 \linewidth]{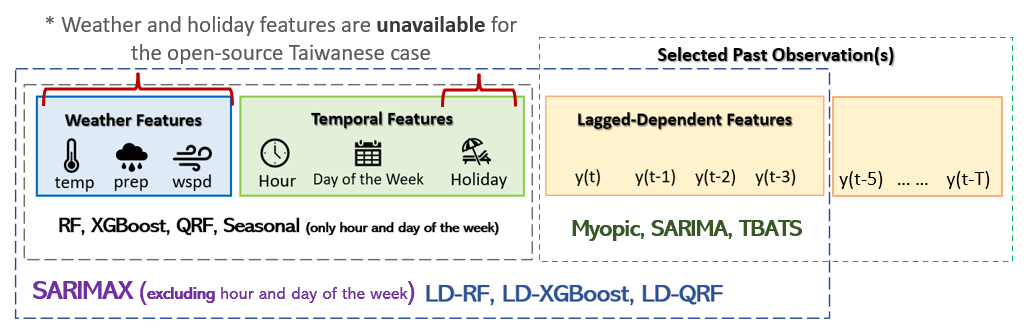}
    \caption{\textcolor{black}{Selected features} for short-term demand forecasting models. Abbreviations: temperature (temp), precipitation (prep), and wind speed (wspd).}
    \label{fig:model_explain}
\end{figure}

Based on the analysis from Section \ref{sec:problem_description}, we propose to utilize three types of input features for the short-term demand forecasting of on-demand meal delivery services. These are the temporal features, weather features, and lagged-dependent features which are the previous observations of demand. 

Meal delivery demand exhibits complex seasonal patterns, as analyzed in Subsection \ref{subset:cases}. These patterns include daily peaks during lunch and dinner hours, higher weekend demand on a weekly scale, and spikes during national holidays. Motivated by these findings, we incorporate temporal features for forecasting, including the day of the week, the hour of the day, and a binary indicator for national holidays. These temporal features are extracted for the specific time interval being predicted. 

Weather conditions affect customers' demand for meal delivery services \cite{liu2021effect,chen2020influence}. In this study, we utilize historical weather data from the open-source climate database of the National Centers for Environmental Information \cite{ncei}. The weather is described using three hourly numerical variables: average temperature (°C), precipitation amount (mm), and wind speed (m/s). It is assumed that weather conditions remain uniform across all zones during the same hour. 

As shown in Figure \ref{fig:model_explain}, the short-term demand forecasting algorithms proposed in Subsection \ref{subsect:forecasting_framework} can be categorized based on the types of input features they employed. Note that the lagged-dependent features for each pick-up zone are extracted from their demand series as explained in Subsection \ref{subset:LD_models}. Benchmarks Myopic, SARIMA, and TBATS rely on historical demand observations only for forecasting. RF, XGBoost, QRF, and Seasonal predictors utilize contextual information including the weather and temporal features. Benchmark SARIMAX jointly considers past demand observations and contextual inputs including weather features and holiday information. Only lagged-dependent models LD-RF, LD-XGBoost, and LD-QRF predict using all three types of features.

We would also like to mention that, since historical date information from the Taiwanese use case is unavailable, only seasonal features (hour and day of the week) and lagged-dependent features are available for demand forecasting. Considering only two seasonal features are available for the none lagged-dependent models in this use case, we propose using historical averaged demand based on the hour and day-of-the-week information as point forecasting benchmark (`Seasonal Average'), and the seasonal approach introduced in Subsection \ref{subsect:forecasting_framework} as the distribution forecasting benchmark.

\subsubsection{Model training and prediction generation}

\textcolor{black}{For the European case,} the original dataset contains orders gathered over 22 weeks. The testing set is derived from the demand series of the final week in the dataset, containing 308 data points to be predicted for each pick-up zone. In reality, platforms often have a limited amount of historical data available for training in some cases, particularly for a fast-expanding platform with new pick-up zones established over time. This challenge can hinder the quality of both univariate time series predictors and ensemble learning-based predictors. \textcolor{black}{To explore the impact of training data size,} we prepare two training datasets of different lengths to fit the demand forecasting models and compare their performance for the European case study. The larger training set contains the first 21 weeks' demand data (6468 data points) for each pick-up zone, while the other training set includes only the demand series from the four weeks (1232 data points) preceding the testing period. \textcolor{black}{Utilizing the date-time details of order transaction data, we can retrieve the historical hourly weather features and generate temporal features. These features are then matched to each data point in the training and testing sets.}

\textcolor{black}{For the Taiwanese use case, the original dataset pre-defines that historical orders from the first 76 days are included in the training set, while orders from the final 14 days are part of the testing set. It results in 7296 and 1344 data points in the training and testing sets respectively.
As in the European case study, we pre-process these orders to create the target demand variable and lagged-dependent input features, along with the hour and day-of-the-week features based on the order placement time. }

In the experiments, zone-specific demand forecasting predictors \textcolor{black}{learn the unique demand pattern processed by each pick-up zone.} 
These models are fitted to \textcolor{black}{the} training set of each pick-up zone. 
The parameter selection for TBATS is done automatically by the algorithm. 
To prevent overfitting of ensemble-learning models, hyperparameters are fine-tuned during the model training phase following a 10-fold cross-validation process.
The list of hyperparameters considered for tuning and their candidate values are detailed in Table \ref{tab:hyper_param} in \ref{appendix:hyperparam}. For SARIMA and SARIMAX modeling, only one seasonal periodicity can be selected due to algorithmic limitations. For \textcolor{black}{computational} complexity concerns, we use a daily seasonal periodicity with 44 \textcolor{black}{and 95 observations} for the European \textcolor{black}{and Taiwanese use cases, respectively} The optimal orders and parameters for SARIMA and SARIMAX are selected via a grid-search process.

Demand predictions in our experiments are generated \textcolor{black}{for each time interval from the testing set} without model retraining for the ensemble learning approaches.
For the point forecasting experiments \textcolor{black}{in the European case study}, we include benchmark models, baseline ensemble-learning models RF, XGBoost, QRF, and their lagged-dependent extensions LD-RF, LD-XGBoost, and LD-QRF. \textcolor{black}{Besides the benchmark models, only lagged-dependent ensemble-learning models LD-RF, LD-XGBoost, and LD-QRF models are included for the Taiwanese use case, due to the unavailability of weather and holiday features.}
The point forecasting outcomes of QRF and LD-QRF are taken as their median predictions. \textcolor{black}{Quantile predictions are generated for quantiles $q = 0.25, 0.5, 0.75$ at each time interval by QRF and LD-QRF to support sequential clustering with distributional forecasted demand insights.}

\textcolor{black}{After completing the short-term demand forecasting step, the predicted demands for all pick-up zones for the current time interval are embedded as dynamic inputs in the sequential clustering procedure. Additionally, the geographical details of these zones are incorporated as static inputs. The clusters are then dynamically generated considering the anticipated short-term demand insights for the current time interval using either the CKMC or CCHC\textcolor{black}{-ICE} clustering algorithms. 
In this study, clustering experiments are conducted only for the European use case, as the pick-up zones in the Taiwanese use case are too large to form local clusters that effectively support micro-delivery operations.} The \textcolor{black}{CKMC algorithm} selects the optimal number of clusters \textcolor{black}{$K_t$} between 3 to 6 based on the highest mean silhouette coefficient, \textcolor{black}{for each time interval}. The implementation of CCHC\textcolor{black}{-ICE} algorithm \textcolor{black}{in the European case study applies the contiguity and example operative constraints defined in subsection \ref{sect:CCHC-ICE}.}
Here, we pre-define the final outcome containing no less than 3 clusters($K_{min}=3$). We also limit each cluster to a maximum of 9 zones ($s_{max} = 9$). The cluster-distance threshold value is defined to be 9 ($D_{max} = 9$).

\subsubsection{Evaluation Scheme}

\textcolor{black}{To evaluate point demand forecasting performance}, we use common metrics including Mean Absolute Error (MAE) and Root Mean Squared Error (RMSE) to assess the magnitude of errors, and Root Mean Squared Logarithmic Error (RMSLE) to measure relative errors. Temporal stability of predictions is analyzed using the standard deviation of residuals (Resid. std.), calculated as the difference between actual and predicted values. For distributional demand forecasting, we report the mean Continuous Ranked Probability Score (MCRPS) across instances. \textcolor{black}{The formulations of MAE, RMSE, RMSLE, and MCRPS are provided in \ref{subset:metrics}.}

For each metric,
the summarized evaluation metric $\Bar{V}$ is computed as the average value of metric $V$ across all pick-up zones, defined as $\bar{V} = \frac{1}{N}\sum^{N}_{i} V^{i}$. The computation of selected metric $V^{i}$ follows the descriptions in \ref{subset:metrics}.
To assess spatial stability, we also report the standard deviation of metric values across pick-up zones. Additionally, to evaluate overfitting, we provide in-sample forecast performance of ensemble learning models using the average MAE.

The predictive cluster insight generated from the short-term predict-then-cluster framework can inform key strategic decisions, such as determining the required fleet size for each cluster or prioritizing fleet relocation to high-demand areas, thereby reducing the need for micro-level management of individual zones. \textcolor{black}{More importantly, these forecasted clusters should reflect the shifting spatial demand dynamics across the city to improve the efficiency of real-time operations.}
Therefore, in this study, we focus on evaluating how well the forward-looking clusters, generated using predicted demand, align with the ideal scenario where clusters are generated based on actual demand. For each pick-up zone, we analyze the extent to which the anticipated demand level with its cluster may differ due to demand forecasting errors introduced to the framework.
Instead of evaluating the quality of clustering outcomes based on downstream operational performance, we measure the perceived estimated demand deviations between the predicted and actual clusters to assess the framework's predictive accuracy.

To perform this evaluation, we first generate the `actual' clusters using the actual demand and geographic information of pick-up zones. 
Next, to identify the actual cluster-level demands, we compute the `actual within-cluster median demand' as the median demand of the cluster it belongs to in the current time window. Similarly, we compute the `predicted within-cluster median demand' using the predicted clusters formed with forecasted demand.  \textcolor{black}{The differences between actual and predicted within-cluster median demand are evaluated using metrics such as MAE, RMSE, RMSLE, and Resid. std.} \textcolor{black}{By comparing these metrics, we can evaluate how closely the predicted clustering outcomes approximate the ideal clustering based on actual demand, providing insights into the predictive clustering framework's effectiveness in real-world applications.}

\subsection{Numerical implementation}\label{subsec:numerical}
\textcolor{black}{Computational efficiency is a key consideration in model selection. On-demand service providers need scalable models to support their extensive service network and favor models that can quickly generate predictive insights within real-time operations.} To analyze the computational efficiency of our proposed algorithms, in the section, we analyze the average model training and result generation time for demand forecasting and clustering algorithms. All the computational experiments are \textcolor{black}{performed} using four concurrent processes on core i7 CPU \textcolor{black}{16 GB of memory} using the same device. 
For each algorithm, we report the training time averaged among the demand predictors for all pick-up zones. 
The one-step prediction generation time is measured by averaging the computation time among the predictions for the whole testing set. Likewise, we report the one-step prediction time as the average among all zone-specific models. \textcolor{black}{An algorithm with low training and prediction generation times is preferred for computational efficiency.} 

During numerical implementation, we discovered that training the SARIMA and SARIMAX models did not converge for the European use case with 21 weeks of training data or the Taiwanese use case's training set within a reasonable computational budget (under an hour of training and formulation tuning per zone-specific model). Thus, SARIMA and SARIMAX are significantly computationally intensive when extensive training data is required for demand forecasting in meal delivery services. Consequently, in the following analysis, we focus on the outcomes from SARIMA and SARIMAX models trained using only 4 weeks of data for the European case.

\begin{table}[h!]
    \centering
    \caption{The \textcolor{black}{model} training and \textcolor{black}{one-step demand} prediction generating time for the experiments in European case study.}
    \resizebox{\textwidth}{!}{%
    % \begin{tabular}{llllllll}
    \begin{tabular}{*{1}{p{3 cm}}|*{2}{p{2.0cm}}*{4}{p{1.3cm}}*{1}{p{1.6cm}}*{1}{p{1.3cm}}*{1}{p{1.5cm}}}
\toprule[1.5pt]
    \textsc{Model} & \textbf{SARIMA} & \textbf{SARIMAX} & \textbf{TBATS} & \textbf{RF} & \textbf{LD-RF} & \textbf{XGB} & \textbf{LD-XGB} & \textbf{QRF} & \textbf{LD-QRF} \\ \hline
    \textsl{Training (4w, sec)} & 1338.8 & 1568.5 & 114.5 & 41.8 & 47.3 & 7.9 & 11.1 & 24.9 & 28.6\\
    \textsl{Training (21w, sec)} & - & - & 324.5 & 104.0 & 155.7 & 22.8 & 35.6 & 90.1 & 117.0\\
    \textsl{Predicting ($10^{-3} sec$)} & 74.2 & 75.6 & 0.4 & 7.5 & 9.1 & 1.4 & 1.9 & 9.3 & 9.5 \\
\bottomrule[1.5pt]
\multicolumn{10}{c}{Abbreviations: 21w (21-week), 4w (4-week), sec (seconds), XGB(XGBoost).}
    \end{tabular}
    }
\label{tab:model_time}
\end{table}
Table \ref{tab:model_time} presents the training and prediction generating times of various demand forecasting algorithms \textcolor{black}{for the European use case}. \textcolor{black}{For the experiments using 4-week training data, SARIMA and SARIMAX models show significantly higher training time than the others.} The training time for XGBoost and LD-XGBoost are the least among the selected forecasting algorithms. \textcolor{black}{As expected,} the training time becomes slightly longer for the ensemble-learning models when lagged dependent variables are included. 
QRF and LD-QRF store observation values directly on regression tree leaves, unlike RF and LD-RF, which apply statistical transformations. In our experiments, we found that this QRF model design slightly reduces training time compared to RF models, but results in significantly larger model file sizes.

\textcolor{black}{For result generation}, all short-term demand forecasting models are capable of producing one-step-ahead predictions within an average computation time of 0.1 seconds. \textcolor{black}{The comparison of computational efficiency for the lagged-dependent predictors (LD-RF, LD-XGBoost, LD-QRF) in the Taiwanese use case is consistent with the European use case.} For the dynamic clustering algorithms, the average computation time for CKMC approach is about 1 second on average, while the average for the proposed CCHC\textcolor{black}{-ICE} algorithm is within 0.1 seconds.

\subsection{Results of short-term point and distributional demand forecasting}\label{subsec:forecasting_result}

\textcolor{black}{This subsection presents the point forecasting performance for selected short-term demand forecasting models.} 

\subsubsection{Point forecasting in the European case study}
The predictive performance of selected models for the European case study is summarized in Table \ref{tab:point}, evaluated using the metrics MAE, RMSE, RMSLE, and Residual standard deviation (Resid. std.). 

Among the benchmark predictors, SARIMA and SARIMAX models trained with the 4-week demand series have obtained the highest prediction accuracy and best performance stability. \textcolor{black}{Notably, the performance of SARIMAX is similar to that of SARIMA, with a slightly higher MAE. This suggests that the additional contextual information in SARIMAX may not have been effectively leveraged to enhance the quality of forecasting.} 
Among the ensemble-learning (EL) models trained with 4-week data, LD-QRF and QRF achieved the lowest MAE and RMSLE respectively using median predictions. LD-QRF also had the smallest average in-sample MAE. Notably, LD-RF excelled with the lowest RMSE and residual variance. 

For models trained with 21-week data, lagged-dependent models performed best in all metrics, with LD-QRF achieving the lowest in-sample MAE, MAE, and RMSLE. While LD-RF had the lowest RMSE, its MAE was higher, suggesting fewer large errors but more small deviations, as supported by the standard deviations. 
Nevertheless, we find that QRF and LD-QRF are less overfitting compared to the other models, measured by the difference between in-sample and out-of-sample MAEs. This resistance to overfitting can be attributed to the inherent robustness of QRF and LD-QRF in handling bias and extreme values within the data. Compared to the myopic benchmark, the LDQRF model trained with 21-week data is capable of reducing average MAE and RMSLE by 26.9\% and 55.0\% respectively. 

Regarding the impact of training data size on models' forecasting performance, EL models perform better with 21-week training data compared to 4-week training data. Specifically, LD-XGBoost showed an 8.5\% improvement in average MAE, while the average RMSLE is reduced by 39.7\% for LD-QRF. Additionally, models trained with 21-week data had reduced residual standard deviations, indicating improved performance stability. Comparing in-sample and out-of-sample MAE, we find that models trained with 21-week data have smaller MAE gaps than those trained with 4-week data. This is expected, as models are prone to noise in a smaller training dataset.  Furthermore, the quality of hyperparameter tuning is improved with more training data, as cross-validation benefits from having more samples per fold.
Overall, the selected EL models outperform the time series benchmarks in both accuracy and performance stability when a large amount of training data is given. 

Accurate demand prediction is crucial for optimizing platform resources and ensuring operational efficiency during peak hours. 
As analyzed in Section \ref{subset:cases}, demand varies by time and day, peaking during dinner hours and weekends. 
Table \ref{tab:point_evening} reports model performance for 48 predictions during weekend dinner periods. Although metric values are higher due to increased demand, the models' relative performance across different metrics still aligns with our findings from the full predictive performance analysis.

\begin{table}[h]
\centering
\caption{\textcolor{black}{European Case Study:} The one-week point forecast results of various models trained with 4-week and 21-week data, the performance is measured by error metrics MAE, RMSE, RMSLE, and standard deviations of residuals (\textcolor{black}{Resid std.}). The standard deviations of the metric value among \textcolor{black}{pick-up zone models} are reported between parentheses. \textcolor{black}{The lowest metrics values in each category are marked in bold with the lowest cross-category values underscored.}}
\label{tab:point_evening}
\resizebox{0.88\textwidth}{!}{%
\begin{tabular}{*{1}{p{2.8cm}}|*{1}{p{2cm}}*{1}{p{1.5cm}}*{4}{p{2cm}}}
\toprule[1.5pt]
 \textsc{Category} & \textsc{Model} & \textbf{In-Sam. MAE} & \textbf{MAE} & \textbf{RMSE}& \textbf{RMSLE} & \textbf{Resid std.}  \\
 \hline
\multirow{4}{*}{Benchmarks  (4W)} & \textbf{Myopic}& - & 1.045 (0.405)&1.554 (0.499)&0.456	 (0.411) &1.554 (0.499)\\
& \textbf{SARIMA}& - & \textbf{0.896} (0.322)&1.255 (0.412)&\textbf{0.324} (0.228)&1.244 (0.407)\\
& \textbf{SARIMAX}& - & 0.915 (0.313)&\textbf{1.246} (0.411)&0.358 (0.225)&\textbf{1.243} (0.408)\\
& \textbf{TBATS}& - & 1.041 (0.368)&1.391 (0.502) &0.585 (0.339)&1.362 (0.453)\\
\hline
\multirow{6}{*}{EL Models  (4W)} & \textbf{RF}& 0.794 & 0.923 (0.340)&1.212 (0.421)&	0.467 (0.350)&1.192 (0.399)\\
& \textbf{XGB}& 0.786 & 0.921 (0.402)&1.238 (0.461)&0.367 (0.299)&1.168 (0.377)\\
& \textbf{QRF}& 0.745 & 0.786 (0.301)&1.229 (0.353)&\textbf{0.239} (0.381)&1.187 (0.352)\\
& \textbf{LD-RF}& 0.790 & 0.873 (0.293)&\textbf{1.168} (0.378)&0.445 (0.312)	& \textbf{1.164} (0.373)\\
& \textbf{LD-XGB}& 0.773 & 0.940 (0.340)&1.250 (0.406)&0.366 (0.267)&1.167 (0.365)\\
& \textbf{LD-QRF}& 0.727 & \textbf{0.779} (0.302)&1.210 (0.351)&0.340 (0.355)&1.180 (0.352)\\
\hline
\multirow{6}{*}{EL Models  (21W)} & \textbf{RF}& 0.785 & 0.863 (0.279)&1.153 (0.361)&0.438 (0.311)&1.147 (0.359)\\
& \textbf{XGB}& 0.819 & 0.887 (0.314)&1.186 (0.393)&0.394 (0.237)&1.140 (0.359)\\
& \textbf{QRF}& 0.757 & 0.768 (0.294)&1.197 (0.339)&0.309 (0.328)&1.169 (0.344)\\
& \textbf{LD-RF}& 0.752 & 0.848 (0.279)&\underline{\textbf{1.141}} (0.354)&0.430 (0.250)&1.139 (0.354)\\
& \textbf{LD-XGB}& 0.815 & 0.860 (0.318)&1.169 (0.396)&0.347 (0.219)&\underline{\textbf{1.128}} (0.342)\\
& \textbf{LD-QRF}& 0.745 & \underline{\textbf{0.764}} (0.297)&1.195 (0.337)&\underline{\textbf{0.205}} (0.298)&1.168 (0.346)\\
\bottomrule[1.5pt]
\multicolumn{7}{l}{(1) Benchmark models SARIMA, SARIMAX, and TBATS are trained with 4-week data;}\\
\multicolumn{7}{l}{(2) Abbreviations: EL (Decision-tree-based ensemble-learning methods), XGB (XGBoost), In-Sam. (In-Sample)}\\
\end{tabular}}
\end{table}

\begin{table}[h]
\centering
\caption{\textcolor{black}{European Case Study:} The one-week point forecast results of various models trained with 4-week and 21-week data, the performance is measured by error metrics MAE, RMSE, RMSLE, and standard deviations of residuals (\textcolor{black}{Resid std.}). The metrics are average among the 48 testing samples corresponding to the dinner times from Friday to Sunday. The standard deviations of the metric value among \textcolor{black}{pick-up zone models} are reported between parentheses. \textcolor{black}{The lowest metrics values in each category are marked in bold with the lowest cross-category values underscored.}}
\label{tab:point}
\resizebox{ 0.85 \textwidth}{!}{%
\begin{tabular}{*{1}{p{2.8cm}}|*{5}{p{2cm}}}
\toprule[1.5pt]
 \textsc{Category} & \textsc{Model} & \textbf{MAE} & \textbf{RMSE}& \textbf{RMSLE} & \textbf{Resid std.}  \\
 \hline
\multirow{4}{*}{Benchmarks  (4W)} & \textbf{Myopic}& 1.693 (0.496)&2.204 (0.624)	&0.618 (0.309) &2.204 (0.624)
\\
& \textbf{SARIMA}& 1.391 (0.454)&\textbf{1.813} (0.575)&\textbf{0.509} (0.320)&1.759 (0.565)
\\
& \textbf{SARIMAX}& \textbf{1.383} (0.457)&\textbf{1.813} (0.587)&0.523 (0.309)&\textbf{1.721} (0.540)
\\
& \textbf{TBATS}& 1.557 (0.673)&2.022 (0.789)	&0.562 (0.427)&1.796 (0.587)
\\
\hline
\multirow{6}{*}{EL Models  (4W)} & \textbf{RF}& 1.342 (0.490) &1.752 (0.611)&0.563 (0.360)&1.669 (0.532)
\\
& \textbf{XGB}& 1.340 (0.461)&1.744 (0.577)&0.587 (0.326)&1.655 (0.505)
\\
& \textbf{QRF}& 1.320 (0.422)&1.776 (0.520)&\textbf{0.402} (0.333)&1.685 (0.462)
\\
& \textbf{LD-RF}& 1.302 (0.435)&\textbf{1.701} (0.547)&0.548 (0.327)&\textbf{1.648} (0.505)
 \\
& \textbf{LD-XGB}& 1.359 (0.447)&1.764 (0.562)&0.568 (0.352)&1.665 (0.488)
 \\
& \textbf{LD-QRF}& \textbf{1.294} (0.456)&1.752 (0.547)&0.426 (0.378)&1.668 (0.491)
\\
\hline
\multirow{6}{*}{EL Models  (21W)} & \textbf{RF}& 1.310 (0.425)&1.689 (0.526)&0.552 (0.347)&1.646 (0.499)
 \\
& \textbf{XGB}& 1.306 (0.409)&1.685 (0.521)&0.552 (0.350)&1.626 (0.473)
\\
& \textbf{QRF}& 1.271 (0.386)&1.718 (0.469)&0.452 (0.348)&1.643 (0.453)
\\
& \textbf{LD-RF}& 1.288 (0.404)&\underline{\textbf{1.662}} (0.500)&0.541 (0.336)&1.634 (0.485)
\\
& \textbf{LD-XGB}& 1.306 (0.414)&1.679 (0.510)&0.549 (0.348)&\underline{\textbf{1.622}} (0.460)
\\
& \textbf{LD-QRF}& \underline{\textbf{1.248}} (0.380)&1.696 (0.463)&\underline{\textbf{0.364}} (0.338)&1.643 (0.442)
\\
\bottomrule[1.5pt]
\multicolumn{6}{l}{(1) Benchmark models SARIMA, SARIMAX, and TBATS are trained with 4-week data;}\\
\multicolumn{6}{l}{(2) Abbreviations: EL (Decision-tree-based ensemble-learning methods), XGB (XGBoost)}
\end{tabular}}
\end{table}

\begin{figure}[h]
    \centering
    \includegraphics[width=\linewidth]{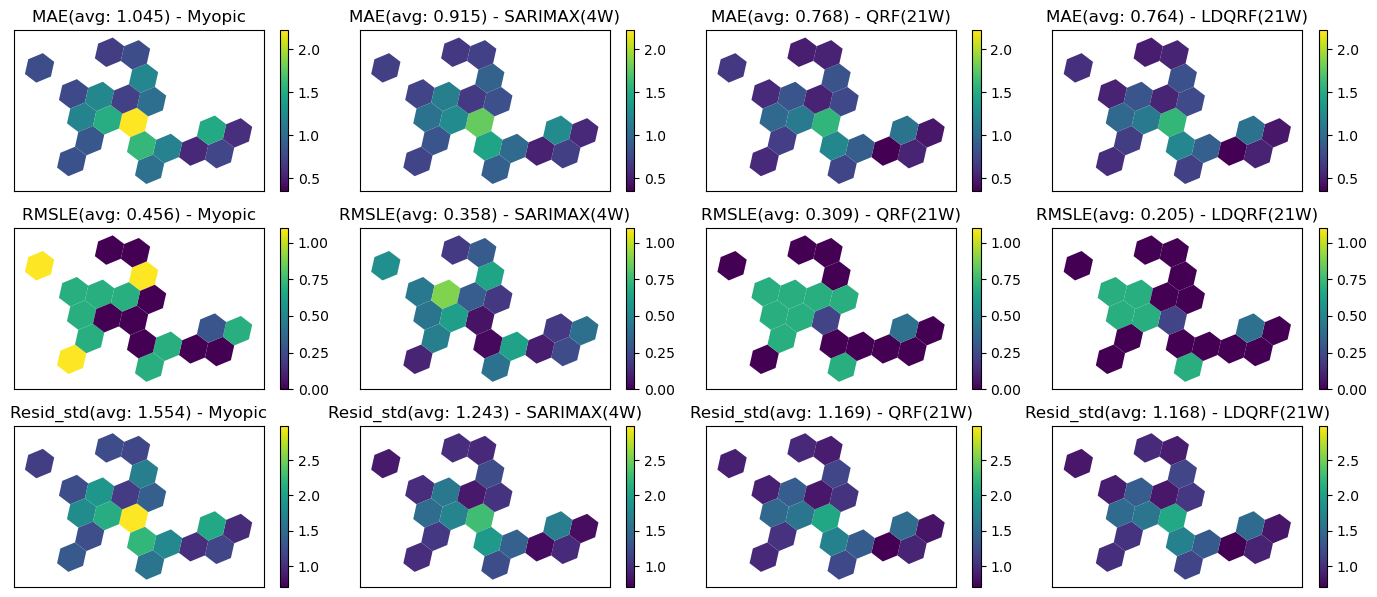}
    \caption{\textcolor{black}{European Case Study:} Heatmap visualization of the pick-up zone-wise metrics MAE, RMSLE, and standard errors of residuals (Resid\_std) for the point predictions generated by Myopic, SARIMAX (with 4-week training data), QRF (with 21-week training data), and LDQRF (with 21-week training data), assessed across all predicted time intervals. \textcolor{black}{The heatmaps employ a color scale to show the relative metrics values, with lighter colors denoting higher values.}}
    \label{fig:heatmap_point}
\end{figure}

\textcolor{black}{Beyond the overall} accuracy summarized from all the pick-up zones, we are also interested in analyzing the spatial variance \textcolor{black}{in} performance to inspect whether a model performs \textcolor{black}{better or worse in particular areas in the city than the others.} 
In Figure \ref{fig:heatmap_point}, we visualize and compare the zone-wise MAE, RMSLE, and standard deviations of residuals from Myopic, SARIMAX, QRF (trained with 21-week data), and LDQRF (trained with 21-week data) models.  
Overall, the metric values are higher for the pick-up zones with higher demand. \textcolor{black}{These zones are located in the city center, which has been previously highlighted} in Figure \ref{Fig:heatmap_15_min_orders}.
Comparing QRF to SARIMA, \textcolor{black}{we observe an} improvement in MAE across the map, indicating higher accuracy in point forecasting overall. However, the RMSLEs of QRF are higher in the center grids compared to SARIMAX's, while the relative errors of QRF are lower at the outer-center areas of the city. 
\textcolor{black}{This implies that QRF slightly struggles in the forecasting for high-demand zones, which may present more variations in short-term demand.}
For the LD-QRF models, there is a significant reduction in RMSLEs in three of the central pick-up zones compared to QRF, \textcolor{black}{highlighting the benefit of incorporating lagged dependencies for capturing demand patterns more accurately in high-demand areas.}

The Mann-Whitney U test is used to determine if there is a significant difference in distribution between groups of samples, while Welch’s t-test assesses the hypothesis of equal means between two populations with unknown variances. 
We applied these tests to compare the residuals \textcolor{black}{(i.e., the difference between actual and predicted demand)} from the models.
With a 95\% confidence interval, most Mann-Whitney U tests reject the hypothesis of no significant distribution differences between the two groups of residuals, except for those obtained from SARIMA and XGBoost (trained with 4-week data). Among EL models, LD-QRF residuals are significantly different from other models when trained with the same data size.
Welch’s t-tests reveal that the residuals from QRF and LD-QRF, whether trained with 4-week or 21-week data, have significantly different mean values compared to Myopic and SARIMA methods. These results indicate that the forecasting improvements with LDQRF are statistically significant.

\begin{figure}[h!]
    \centering
    \includegraphics[width=0.47\linewidth]{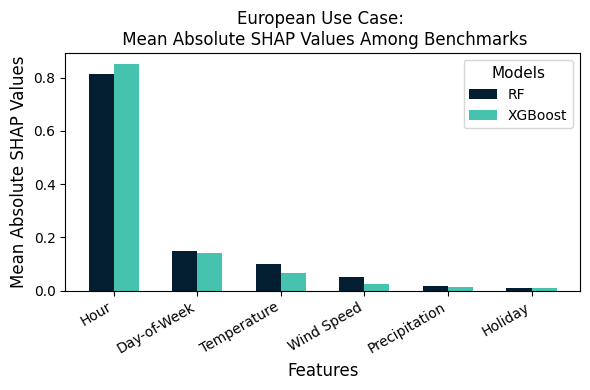}
    \includegraphics[width=0.48\linewidth]{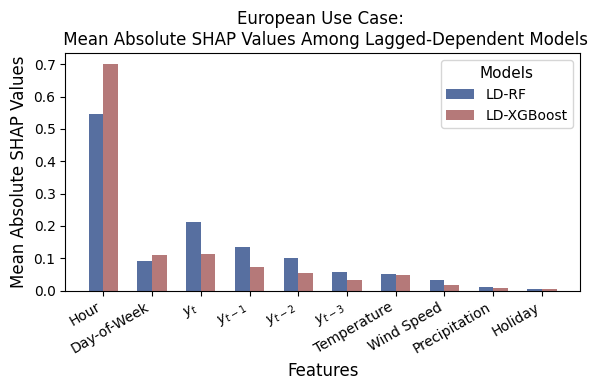}
    \caption{\textcolor{black}{European Case Study:} Mean absolute SHAP values from the feature importance analysis. The left sub-figure compares feature importance among benchmark models (RF and XGBoost). The right sub-figure illustrates feature importance for lagged-dependent models (LD-RF and LD-XGBoost), which incorporate lagged demand features ($y_t$, $y_{t-1}$, $y_{t-2}$, and $y_{t-3}$). All selected models are trained with 21-week data.}
    \label{fig:SHAP}
\end{figure}

\textcolor{black}{To enhance the interpretability of the trained short-term demand forecasting models, we perform a post-hoc analysis using SHapley Additive exPlanations (SHAP)\cite{lundberg2017unified}, to quantify feature importance. SHAP values provide a unified measure of each feature's contribution to the model's predictions, evaluated over the training instances. A feature is considered important if its mean absolute SHAP value is comparatively higher. These values are non-negative, with larger values indicating greater impact on prediction outcomes.}
\textcolor{black}{Given that SHAP has been optimally designed for classic tree models like Random Forest (RF) and XGBoost \cite{lundberg2017unified}, we calculate the absolute SHAP values for these tree-based ensemble learning models, including their lagged-dependent variants (LD-RF and LD-XGBoost). The selected models are trained with 21-week data.
The mean absolute SHAP values are evaluated for each zonal demand predictor across 20 pick-up zones in the European use case.} 

\textcolor{black}{Figure \ref{fig:SHAP} compares the mean absolute SHAP values of the features averaged over the all zonal predictors for RF and XGBoost models on the left sub-figure, and evaluates those for LD-RF and LD-XGBoost on its right sub-figure. For all models, hour-of-the-day feature has the highest importance among all, meaning the daily demand pattern effects dominate in forecasting modeling. In contrast, weather and holiday features are seen less influential by the models. In the SHAP analysis for lagged-dependent models, lagged demand features emerge as the second most important contributors to predictions. While the temporal features still dominates the predictions made by LD-XGBoost, LD-RF distributes greater importance to the lagged demand features. When connected to the forecasting results in Table \ref{tab:point}, where LD-RF outperforms LD-XGBoost in terms of MAE, RMSE, and RMSLE when trained on 21-week data, this indicates that LD-RF's more refined utilization of lagged demand features enhances its predictive performance.}

\subsubsection{Point forecasting in Taiwanese case study}

\begin{table}[h]
\centering
\caption{\textcolor{black}{Taiwanese Case Study: The point forecast results of benchmark and lagged-dependent ensemble-learning (LDEL) models of the provided 14-day testing data. All models are trained with the provided 76-day training data. The forecasting performance is measured by error metrics MAE, RMSE, RMSLE, and standard deviations of residuals (Resid std.). The standard deviations of the metric value among pick-up zone models are reported in parentheses.} \textcolor{black}{The lowest metrics values in each category are marked in bold with the lowest cross-category values underscored.}}
\label{tab:point_tw}
\resizebox{0.9\textwidth}{!}{%
\begin{tabular}{*{1}{p{2.3cm}}|*{1}{p{2.6cm}}*{1}{p{1.5cm}}*{4}{p{2cm}}}
\toprule[1.5pt]
 \textsc{Category} & \textsc{Model} & \textbf{In-Sam. MAE} & \textbf{MAE} & \textbf{RMSE} & \textbf{RMSLE} & \textbf{Resid std.}  \\
 \hline
\multirow{3}{*}{Benchmarks} & \textbf{Myopic} & - & 3.192 (1.439) & 4.567 (1.969) & 0.458 (0.323) & 4.567 (1.969) \\
& \textbf{TBATS} & - & 2.896 (1.825) & \textbf{3.980} (2.330) & 0.360 (0.315) & \textbf{3.779} (1.960) \\
& \textbf{Seasonal Avg.} & - & \textbf{2.819} (1.474) & 4.187 (2.183) & \textbf{0.353} (0.335) & 3.843 (1.880) \\
\hline
\multirow{3}{*}{LDEL Models} & \textbf{LD-RF} & 1.771 & 2.551 (1.167) & 3.696 (1.672) & 0.368 (0.358) & 3.645 (1.636) \\
& \textbf{LD-XGB} & 2.126  & \underline{\textbf{2.490}} (1.134) & \underline{\textbf{3.593}} (1.614) & 0.352 (0.327) & \underline{\textbf{3.537}} (1.576) \\
& \textbf{LD-QRF} & 1.879  & 2.558 (1.169) & 3.709 (1.680) & \underline{\textbf{0.346}} (0.359) & 3.658 (1.646) \\
\bottomrule[1.5pt]
\multicolumn{7}{l}{(1) Benchmark model TBATS is trained with 4-week data;}\\
\multicolumn{7}{l}{(2) Abbreviations: LDEL (Lagged-dependent decision-tree-based ensemble-learning methods), XGB (XGBoost),}\\
\multicolumn{7}{l}{Avg. (Average), In-Sam. (In-Sample)}\\
\end{tabular}}
\end{table}
\textcolor{black}{The short-term demand point forecasting results for the Taiwanese use case are presented in Table \ref{tab:point_tw}. Given historical date information is unavailable for the Taiwanese use case, only hour and day-of-the-week features are available, along with the lagged-dependent information extracted from the demand series. Therefore, we include only lagged-dependent ensemble-learning approaches for comparison against the benchmarks.}

\textcolor{black}{The values of MAE, RMSE, and Residual Standard Deviation are higher in the Taiwanese use case compared to the European case, primarily due to higher average demand per 15-minute interval and the limited availability of contextual information for the ensemble learning models. Among the benchmarks, the TBATS and Seasonal Average models exhibit similar forecasting accuracy, both outperforming the Myopic predictor. Notably, all lagged-dependent ensemble learning models outperform the benchmarks in terms of MAE, RMSE, and Resid. std., with LD-XGBoost and LD-QRF achieving the highest overall prediction accuracy.} 

\begin{figure}[h!]
    \centering
    \includegraphics[width=\linewidth]{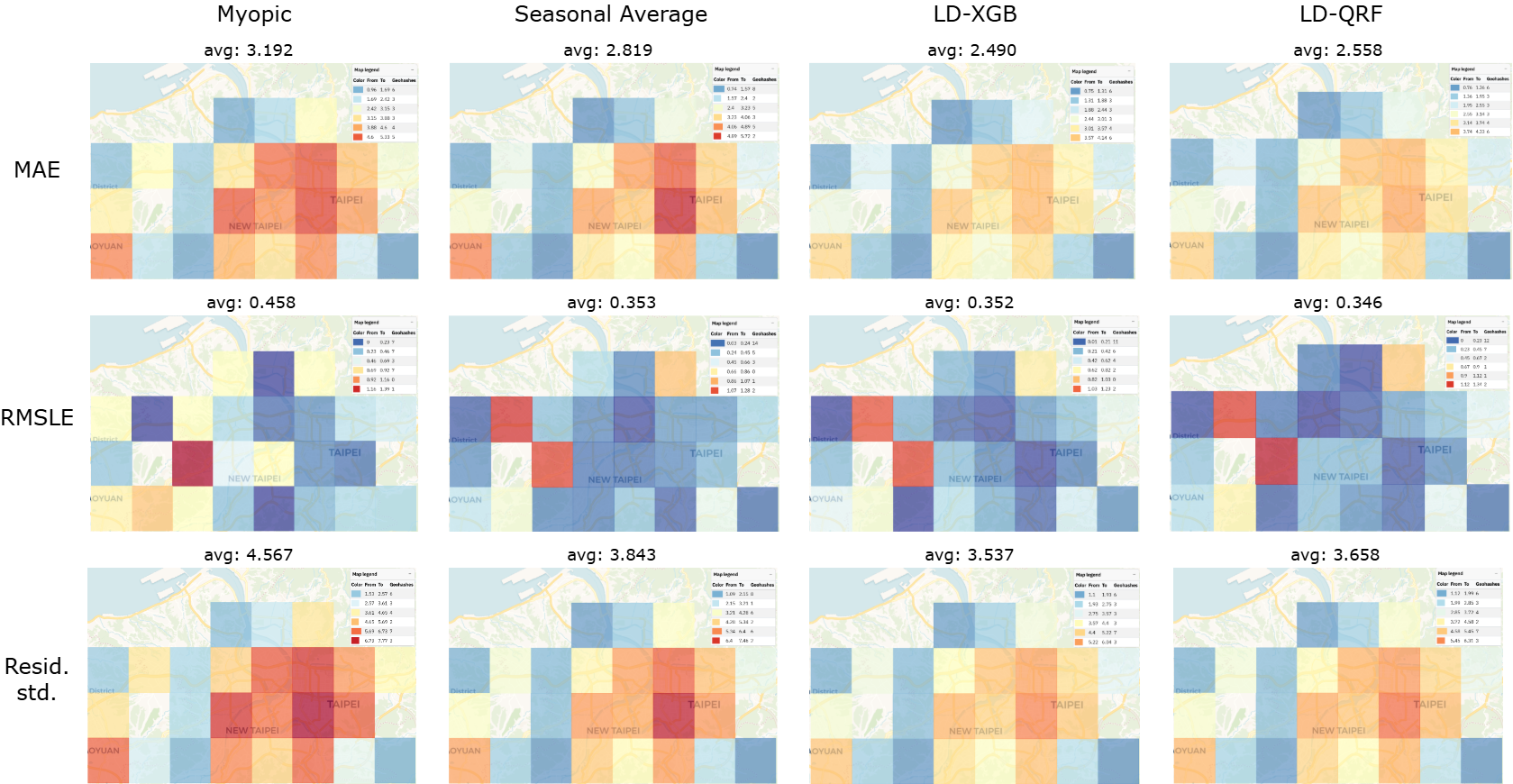}
    \caption{\textcolor{black}{Taiwanese Case Study: Heatmap visualization of the pick-up zone-wise metrics MAE, RMSLE, and standard errors of residuals (Resid. std.) for the point predictions generated by benchmarks Myopic, Seasonal Average models, lagged-dependent ensemble learning predictors LD-XGBoost and LD-QRF, assessed across all predicted time intervals. The heatmaps employ a color scale to show the relative metrics values, with hotter (closer to red) colors denoting higher values.}}
    \label{fig:error_heatmap_taiwan}
\end{figure}

\textcolor{black}{Similar to Figure \ref{fig:heatmap_point} for the European case study, Figure \ref{fig:error_heatmap_taiwan} presents a comparative analysis of forecasting errors for pick-up demand across zones in a Taiwanese city. The heatmaps visualize three error metrics (MAE, RMSLE, and Resid. std.) for four selected models Myopic, Seasonal Average, LD-XGBoost, and LD-QRF. 
Relating to Figure \ref{fig:heatmap_TW}, the lagged-dependent models (LD-XGBoost and LD-QRF) consistently reduce MAE and Resid. std. across high-demand zones compared to the benchmarks, particularly Myopic and Seasonal Average. This improvement indicates that the inclusion of lagged features enhances prediction accuracy for high-demand areas. 
However, for RMSLE, the highest error values are observed in two moderate-demand zones on the left side of the map. This suggests that demand patterns of these zones are less predictable, when only temporal and lagged demand features are considered. Notably, the Myopic model yields lower RMSLE values for one of these zones, implying that simpler historical averages may better capture its demand dynamics compared to the more complex lagged-dependent methods.}

As in the European case study, we performed statistical tests on the prediction residuals to compare the significance of performance differences between predictors. The outcomes of the Mann-Whitney U test and Welch’s t-test suggest that the superior performance of LDEL models over benchmark models is statistically significant at a 95\% confidence interval.

\begin{figure}[h!]
    \centering
    \includegraphics[width=0.5\linewidth]{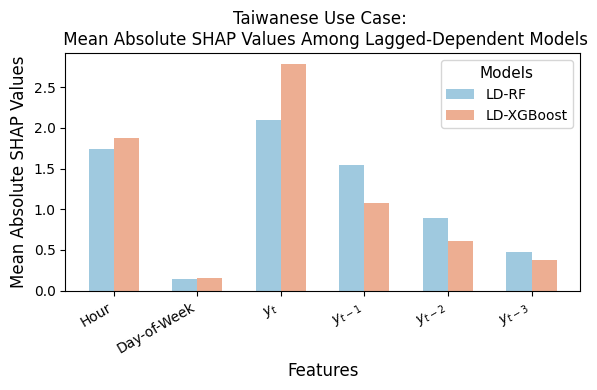}
    \caption{\textcolor{black}{Taiwanese Case Study: Mean absolute SHAP values from the feature importance analysis for LD-RF and LD-XGBoost.}}
    \label{fig:SHAP_TW}
\end{figure}

\textcolor{black}{Mean absolute SHAP values are computed to analyze the feature importance of temporal and lagged demand features in the Taiwanese case study. Averaged across 25 zonal models within the Taiwanese meal delivery service network, lagged demand features exhibit diminishing importance, as shown in Figure \ref{fig:SHAP_TW}, with $y_t$ being the most critical predictor, followed by $y_{t-1}$, $y_{t-2}$, and $y_{t-3}$.
Compared to the feature importance findings from the European use case (presented in Figure \ref{fig:SHAP}), lagged demand features are assigned greater importance in the Taiwanese use case, likely due to the absence of weather and holiday features in the dataset. This finding suggests that, in the absence of external contextual features, the model compensates by relying more heavily on historical demand patterns, highlighting the predictive contributions of the lagged demand variables.
}

\subsubsection{Distributional forecasting performance}

\textcolor{black}{To investigate the distributional forecasting capability of the proposed short-term distributional demand predictors QRF and its lagged-dependent extension LD-QRF, 
we estimate their MCRPS values and compare them with the Seasonal distributional forecasting benchmark. If QRF and LD-QRF models more accurately capture the underlying demand distribution, utilizing weather conditions and sequential fluctuations in demand, they will exhibit lower MCRPS values compared to the Seasonal benchmark. } 
\begin{table}[ht]
    \centering
    \caption{\textcolor{black}{The Mean Continuous Ranked Probability Score (MCRPS) of probabilistic predictions given by benchmark Seasonal method, QRF, and LD-QRF models for both European and Taiwanese case studies.}}
    \resizebox{0.5\textwidth}{!}{%
    \begin{tabular}{lcccc}
        \toprule[1 pt]
        \textsc{Category} & \textsc{Model} & \textsc{Training} & \textsc{Testing} \\
        \midrule
        \textit{European Use Case} & Seasonal & 0.513 (0.201) & 0.571 (0.220) \\
        \textit{4-week training data}& QRF & 0.484 (0.183) & 0.564 (0.215) \\
        & LD-QRF & \textbf{0.423 (0.140)} & \textbf{0.560 (0.214)} \\
        \midrule
        \textit{European Use Case} & Seasonal & 0.559 (0.216) & \textbf{0.554 (0.202)} \\
        \textit{21-week training data}& QRF & 0.515 (0.199) & 0.554 (0.205) \\
        & LD-QRF & \textbf{0.456 (0.170)} & 0.555 (0.207) \\
        \midrule
        \multirow{2}{*}{\textit{Taiwanese Use Case}} & Seasonal & 1.416 (0.563) & 1.728 (0.742) \\
        & LD-QRF & \textbf{0.931 (0.275)} & \textbf{1.430 (0.504)} \\
        \bottomrule[1 pt]
    \end{tabular}}
    \label{tab:MCRPS}

\end{table}

\textcolor{black}{The evaluation results for distributional forecasting performance are presented in Table \ref{tab:MCRPS}.
For both European and Taiwanese use cases, LD-QRF demonstrates the best fit to the historical demand distribution within the training sets among the distributional predictors. 
In the European case, when 21-week training data is available, all three models demonstrate similar distributional forecasting performance over the testing set. The distributional forecasting quality of QRF and LD-QRF is improved by using a more extensive training dataset, especially for the high-demand pick-up zones.
Across both training durations (4 weeks and 21 weeks), LD-QRF outperforms other models on both the training and testing datasets.
The superior performance of LD-QRF predictors is more pronounced in the Taiwanese case, where demand patterns exhibit higher values and larger variations. Compared to the Seasonal benchmark, the average MCRPS for LD-QRF on the training set is reduced by 34.2\% (from 1.416 to 0.931) and on the testing set by 17.2\% (from 1.728 to 1.430).
These results highlight LD-QRF’s ability to effectively learn underlying demand distributions, even when training data is limited. The model's advantage in distributional forecasting becomes more significant as demand series exhibit higher magnitudes and greater variability.}

\subsection{\textcolor{black}{Performance of short-term predict-then-cluster in the European case study}} \label{subsec:clustering_result}

From the previous experiment, we have found that the lagged-dependent ensemble-learning models are able to generate accurate point predictions. In the \textcolor{black}{dynamic predict-then-cluster framework}, we combine the geographical information and the \textcolor{black}{predicted demand information} of pick-up zones as inputs to the proposed \textcolor{black}{Constrained K-Means Clustering (CKMC) and Contiguity Constrained Hierarchical Clustering with Iterative Constraint Enforcement (CCHC\textcolor{black}{-ICE}) approaches to generated predicted demand hot- and coldspot insights of the service network}.
\textcolor{black}{To evaluate how the quality of demand predictions influences the outcomes of each clustering approach, we measure the difference between predicted and actual within-cluster median demand value for each pick-up zone, as detailed in Subsection \ref{subsec:experiment_design}. In our experiments for the predict-then-cluster framework, we use point demand predictions from forecasting algorithms SARIMA, SARIMAX, QRF, LD-RF, LD-XGBoost, and LD-QRF. These predictive models have shown competitive short-term forecasting performance. Additionally, we use quantile predictions from QRF and LD-QRF to provide insights on demand uncertainties for the clustering approaches.} 

\subsubsection{\textcolor{black}{Short-term predict-then-cluster with CKMC}}

\begin{table}[h]
\centering
\caption{\textcolor{black}{European Case Study:} The one-week \textcolor{black}{dynamic clustering results of \textbf{Constrained K-Means Clustering (CKMC)} using predicted demand information from various forecasting models}, the performance is measured by the difference between the `actual within-cluster median demand' and `predicted within-cluster median demand', using error metrics MAE, RMSE, RMSLE, and standard deviations of residuals (\textcolor{black}{Resid std.}). The standard deviations of the metric value among \textcolor{black}{pick-up zone models} are reported between parentheses. \textcolor{black}{The lowest metrics values in each category are marked in bold with the lowest cross-category values underscored.}}
\label{tab:hotspot}
\resizebox{0.85\textwidth}{!}{%
\begin{tabular}{*{1}{p{2.6cm}}|*{1}{p{2.6cm}}*{5}{p{2cm}}}
\toprule[1.5pt]
 \textsc{Category} & \textsc{Model} & \textbf{MAE} & \textbf{RMSE}& \textbf{RMSLE} & \textbf{Resid std.}  \\
 \hline
\multirow{2}{*}{Benchmarks  (4W)} 
& \textbf{SARIMA}& \textbf{0.685}(0.256)	& \textbf{0.934}(0.361)	&0.395(0.065)&\textbf{0.919}(0.351)
\\
& \textbf{SARIMAX}& 0.721(0.240)&0.940(0.338)&0.414(0.044)&0.923(0.341)
\\
\hline
\multirow{6}{*}{EL Models  (4W)} & \textbf{QRF}& \textbf{0.505}(0.233)&\textbf{0.825}(0.287)&0.171(0.271)&0.804(0.284) \\
& \textbf{LD-RF}& 0.658(0.189)&0.830(0.278)	&0.457(0.179)&\textbf{0.798}(0.298)
\\
& \textbf{LD-XGB}& 0.722(0.239)	&0.923(0.316)	&0.301(0.257)	&0.856(0.276)
\\
& \textbf{LD-QRF}& 0.533(0.257)	&0.871(0.297)	&\textbf{0.171}(0.240)	&0.839(0.282)
\\
& \textbf{Quan. QRF}& 0.509(0.244)	&0.834(0.301)	&0.171(0.323)	&0.813(0.292)
\\
& \textbf{Quan. LD-QRF}& 0.541(0.258)	&0.853(0.305)	&0.137(0.327)	&0.826(0.281)
\\
\hline
\multirow{6}{*}{EL Models  (21W)} 
& \textbf{QRF}& 0.500(0.219)&0.801(0.263)&\underline{\textbf{0.102}}(0.169)&0.789(0.257)
\\
& \textbf{LD-RF}& 0.620(0.163)&0.792(0.248)&0.422(0.203)&0.758(0.266)
\\
& \textbf{LD-XGB}& 0.636(0.241)&0.832(0.322)	&0.282(0.207)&0.776(0.279)
\\
& \textbf{LD-QRF}& 0.502(0.228)&0.816(0.277)	&0.136(0.211)&0.805(0.271)
\\
& \textbf{Quan. QRF}& 0.477(0.234) & \underline{\textbf{0.791}}(0.281)	&0.171(0.323)	& \underline{\textbf{0.774}}(0.264)
\\
& \textbf{Quan. LD-QRF}& \underline{\textbf{0.475}}(0.227)	& 0.791(0.289)	&0.136(0.274)	&0.774(0.276)
\\
\bottomrule[1.5pt]
\multicolumn{6}{l}{(1) Benchmark models SARIMA and SARIMAX are trained with 4-week data;}\\
\multicolumn{6}{l}{(2) Abbreviations: EL (Decision-tree-based ensemble-learning methods), XGB (XGBoost), Quan (Quantile)}\\
\end{tabular}}
\end{table}

Table \ref{tab:hotspot} presents the dynamic clustering performance using CKMC, evaluated by the differences between actual and predicted within-cluster median demands, using metrics such as MAE, RMSE, RMSLE, and the standard deviation of residuals. Among the benchmark methods, SARIMA achieves the highest hotspot forecasting accuracy based on point predictions.
For ensemble-learning models trained on 4-week data, QRF using median predictions \textcolor{black}{attains the lowest average MAE, RMSE, and RMSLE}. In the demand point forecasting experiments, QRF has obtained the least relative error measured by RMSLE (Table \ref{tab:point}). \textcolor{black}{This suggests that higher forecasting accuracy for high-demand pick-up zones helps generate clusters that closely match those based on actual demand.} 

Applying predictions generated by models trained on 21-week data, \textcolor{black}{we observe lower average metric values compared to those on 4-week data.} However, the differences in MAE and RMSE between the two training durations are relatively modest, suggesting that high-quality dynamic cluster insights can still be generated with limited training data during the forecasting step.
The lowest MAE and RMSE are achieved using distributional predictions from QRF and LD-QRF trained by 21-week data. This indicates that considering demand uncertainty through quantile predictions \textcolor{black}{enhances clustering quality and performance stability}.
Furthermore, clustering quality has been improved by applying the quantile predictions from QRF and LD-QRF trained with 21-week data, compared to the case with 4-week data.  These results align with the distributional prediction performance analysis, suggesting that accurately describing future demand uncertainties in high-demand zones is crucial for generating high-quality clusters.
According to this hotspot forecasting, using quantile predictions from an LD-QRF model trained with 21-week data has reduced the MAE by 30.66\%, RMSE by 15.30\%, and RMSLE by 65.57\% respectively compared to the SARIMA benchmark. 

\subsubsection{\textcolor{black}{Short-term predict-then-cluster with CCHC-ICE}}

\begin{table}[!h]
\centering
\caption{\textcolor{black}{European Case Study:} \textcolor{black}{The one-week dynamic clustering results of \textbf{Contiguity Constrained Hierarchical Clustering with Iterative Constraint Enforcement (CCHC\textcolor{black}{-ICE})} using predicted demand information from various forecasting models, the performance is measured by the difference between the `actual within-cluster median demand' and `predicted within-cluster median demand', using error metrics MAE, RMSE, RMSLE, and standard deviations of residuals (Resid std.). The standard deviations of the metric value among models of pick-up zones are reported between parentheses.} \textcolor{black}{The lowest metrics values in each category are marked in bold with the lowest cross-category values underscored.}}
\label{tab:cchc_performance}
\resizebox{0.85\textwidth}{!}{%
\begin{tabular}{*{1}{p{2.6cm}}|*{1}{p{2.6cm}}*{4}{p{2cm}}}
\toprule[1.5pt]
 \textsc{Category} & \textsc{Model} & \textbf{MAE} & \textbf{RMSE}& \textbf{RMSLE} & \textbf{Resid std.}  \\
 \hline
\multirow{2}{*}{Benchmarks (4W)} 
& \textbf{SARIMA} & \textbf{0.835} (0.348) & \textbf{1.187} (0.460) & \textbf{0.391} (0.214) & \textbf{1.179} (0.455) \\
& \textbf{SARIMAX} & 0.862 (0.331) & 1.188 (0.450) & 0.439 (0.204) & 1.183 (0.449) \\
\hline
\multirow{6}{*}{EL Models (4W)} 
& \textbf{QRF} & 0.754 (0.333) & 1.175 (0.392) & \textbf{0.366} (0.416) & 1.146 (0.398) \\
& \textbf{LD-RF} & 0.816 (0.305) & 1.123 (0.408) & 0.531 (0.169) & 1.114 (0.408) \\
& \textbf{LD-XGB} & 0.806 (0.300) & \textbf{1.114} (0.407) & 0.464 (0.177) & \textbf{1.104} (0.404) \\
& \textbf{LD-QRF} & \textbf{0.740} (0.344) & 1.187 (0.408) & 0.347 (0.370) & 1.152 (0.409) \\
& \textbf{Quan. QRF} & 0.760 (0.333) & 1.190 (0.403) & 0.347 (0.370) & 1.159 (0.401) \\
& \textbf{Quan. LD-QRF} & \textbf{0.740} (0.340) & 1.180 (0.409) & 0.465 (0.317) & 1.149 (0.412) \\
\hline
\multirow{6}{*}{EL Models (21W)} 
& \textbf{QRF} & 0.715 (0.332) & 1.149 (0.384) & 0.511 (0.287) & 1.127 (0.389) \\
& \textbf{LD-RF} & 0.793 (0.310) & 1.095 (0.395) & 0.537 (0.167) & 1.086 (0.399) \\
& \textbf{LD-XGB} & 0.783 (0.299) & \underline{\textbf{1.086}} (0.390) & 0.399 (0.167) & \underline{\textbf{1.078}} (0.394) \\
& \textbf{LD-QRF} & 0.720 (0.340) & 1.145 (0.403) & 0.381 (0.368) & 1.123 (0.403) \\
& \textbf{Quan. QRF} & \underline{\textbf{0.705}} (0.342) & 1.142 (0.408) & 0.392 (0.359) & 1.119 (0.409) \\
& \textbf{Quan. LD-QRF} & 0.714 (0.340) & 1.133 (0.402) & \underline{\textbf{0.361}} (0.338) & 1.113 (0.402) \\
\bottomrule[1.5pt]
\multicolumn{6}{l}{(1) Benchmark models SARIMA and SARIMAX are trained with 4-week data;}\\
\multicolumn{6}{l}{(2) Abbreviations: EL (Decision-tree-based ensemble-learning methods), XGB (XGBoost), Quan. (Quantile)} \\
\end{tabular}}
\end{table}

Table \ref{tab:cchc_performance} presents the evaluation of clustering performance using CCHC\textcolor{black}{-ICE} approach. 
Compared to CKMC, the application of CCHC\textcolor{black}{-ICE} results in higher metric values due to the additional constraints imposed by contiguity and operational requirements during cluster formation. 
Variations in actual versus predicted demand data can significantly influence the formation of clusters throughout the hierarchical process.
The CCHC-ICE outcomes show improved performance when using EL demand predictors within the predict-then-cluster framework, yielding lower metric values compared to SARIMA or SARIMAX predictors. This improvement is more pronounced when EL models are trained on 21-week data, as extended training data enhances the clustering accuracy.
Overall, clustering outcomes generated using distributional predictions align more closely with clusters formed using actual demand. 
Among the tested approaches, quantile predictions from QRF and LD-QRF trained on 21-week data achieve the best results: QRF delivers the lowest average MAE, while LD-QRF achieves the lowest RMSLE. However, with limited training data (4 weeks), the less precise quantile predictions from QRF and LD-QRF may lead to reduced clustering performance, as demand uncertainty is not fully captured.
Interestingly, the lowest RMSE and Resid. std. are obtained using LD-XGBoost as the point demand predictor. However, its slightly higher MAE and RMSLE suggest that LD-XGBoost may be less optimal for zones with higher demand variability, where dynamic shifts in cluster assignments occur over time. 
\textcolor{black}{This observation aligns with the feature importance analysis (Figure \ref{fig:SHAP}), which shows that lagged demand features are assigned slightly lower importance in LD-XGBoost models compared to LD-RF models.}

\begin{figure}[h]
    \centering
    \includegraphics[width= 0.8\linewidth]{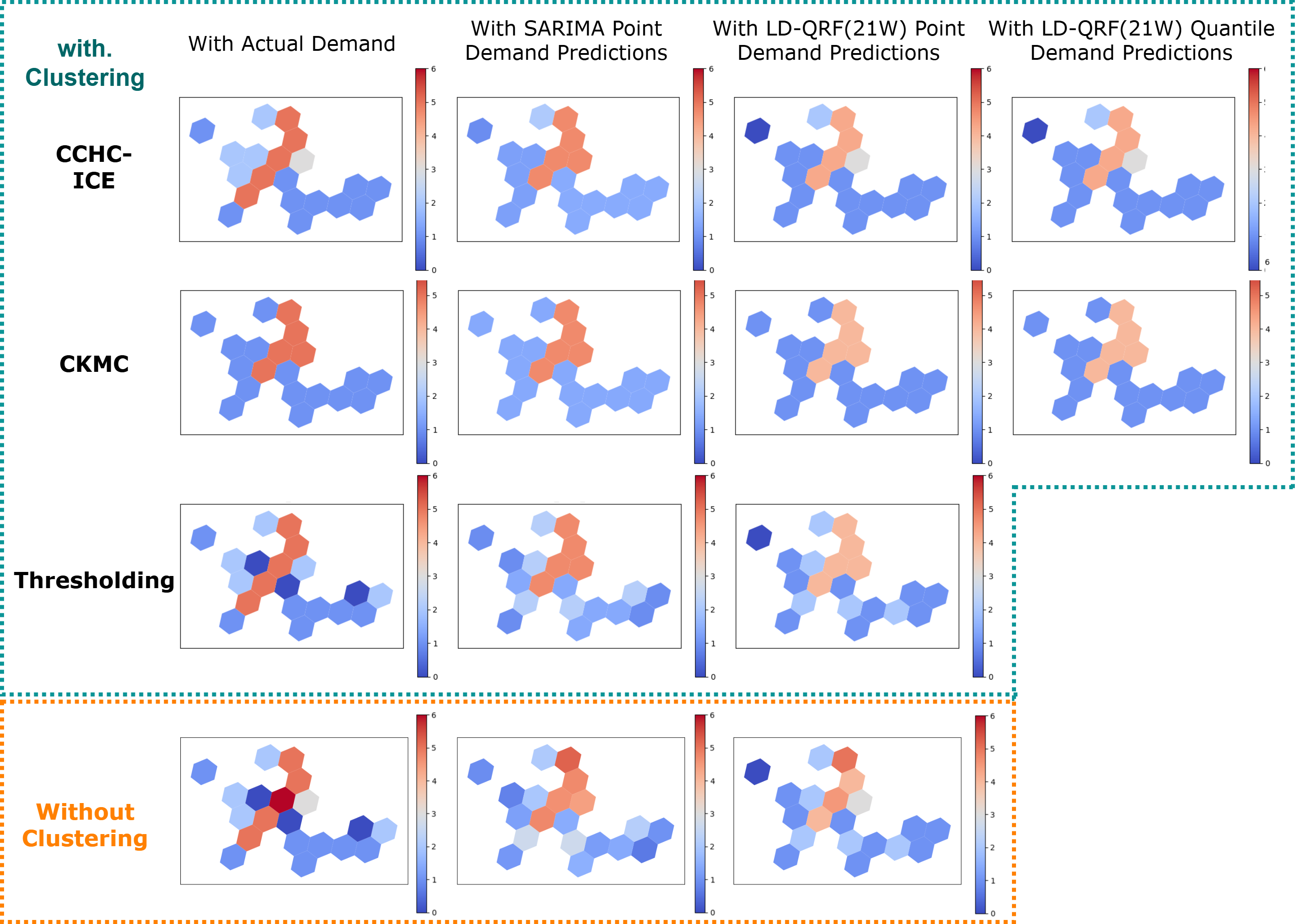}
    \caption{\textcolor{black}{European Case Study:} Heatmap visualization of the actual and predicted pick-up zonal demand for 19:00-19:15 on Monday, September $14^{th}$ and the resulting clusters with these demand estimates as inputs. Clusters are derived from three different approaches: Thresholding (ignoring location), CKMC (considering geographical proximity), and CCHC\textcolor{black}{-ICE} (enforcing geographical contiguity). Demand inputs for clustering include actual demand, SARIMA point predictions, and LD-QRF median and quantile predictions (trained with 21 weeks of data) for CKMC and CCHC\textcolor{black}{-ICE} methods. The color scale represents demand density, with warmer colors indicating higher order volumes.}
    \label{fig:heatmap_cluster}
\end{figure}

\subsubsection{\textcolor{black}{Example visualization of the short-term predict-then-cluster outcomes}}

Figure \ref{fig:heatmap_cluster} presents examples of visualized outcomes from three representative dynamic clustering approaches (CCHC\textcolor{black}{-ICE}, CKMC, and Thresholding) applied to the interval from 19:00 to 19:15 on Monday, September $14^{th}$ \textcolor{black}{for the European case study}. Clustering is performed using either the actual demand or demand estimations of the pick-up zones as inputs.
The demand estimates include point predictions from SARIMA, LD-QRF, and quantile predictions from LD-QRF (trained with 21 weeks of data). \textcolor{black}{Additionally, the figure includes the resulting demand heatmaps without clustering, providing a direct comparison to evaluate the impact of clustering.}
In the heatmaps, the color of each hexagon represents the within-cluster median actual or predicted demand of the cluster to which each pick-up zone belongs. Therefore, all zones within the same cluster are represented by the same color. Cooler colors (closer to blue) indicate lower within-cluster demand, while hotter colors (closer to red) indicate higher demand. Similar median values between clusters may result in similar colors across different clusters.

\textcolor{black}{The heatmaps in the bottom row of Figure \ref{fig:heatmap_cluster} show the zonal demand information without clustering, where each grid is colored based on its individual demand level. In this example, the heatmaps generated with demand estimates provide a rough identification of grids with comparatively higher future demand levels.}

Thresholding is a straightforward clustering approach that categorizes pick-up zones based solely on demand, without considering geographical proximity. In this approach, clusters are formed by grouping zones according to the percentiles of predicted or actual demand values. As showed in Figure \ref{fig:heatmap_cluster}, for each heatmap visualized for clusters under thresholding approach, four clusters are formed, representing zones with demand levels in the highest 75\% of demand, 50-75\% demand, 25-50\% demand, and the lowest 25\% of demand. In the example, thresholding effectively identifies the highest demand cluster near the city center using demand predictions.

CKMC generates clusters by considering geographic proximity and constraining the minimum cluster size to three. The optimal number of clusters, ranging from 2 to 6, is automatically selected based on the highest mean silhouette score for each clustering iteration.
The resulting clusters are visualized in the second row of heatmaps in Figure \ref{fig:heatmap_cluster}. 
Using regardless the actual or predicted zonal demand as inputs, each heatmap is divided into the same two major clusters, highlighting the city's high- and low-demand regions. Notably, the minimum cluster size constraint does not always benefit the creation of cohesive service neighborhoods. \textcolor{black}{Compared to the case without clustering,} some zones within the same cluster are geographically distant from each other and could be more appropriately separated into different clusters.

Taking the geographical adjacency relations of the pick-up zones as inputs, CCHC\textcolor{black}{-ICE} generates clusters based on demand similarity while strictly adhering to geographical contiguity \textcolor{black}{and other cluster configuration requirements defined by the platform}.
In the case study, at least 3 clusters and a maximum number of 9 zones are defined as constraints for CCHC\textcolor{black}{-ICE}. 
Utilizing the actual demand as input, CCHC\textcolor{black}{-ICE} identifies a high-demand cluster at the center, with median-demand clusters to the left and right of it. 
Among the clustering outcomes with predicted demand as input, the results using predictions from LD-QRF also correctly identify the median-demand cluster on the right (shown as the grid colored in gray). 
\textcolor{black}{Compared to CKMC and thresholding methods with the same demand estimates as input, the CCHC-ICE outputs are more informative, as they highlight the relatively lower predicted demand level of this pick-up zone to its surrounding high-demand regions.}

\section{Discussion and Future Research}
\label{sec:disc}
\textcolor{black}{This section discusses the practical implications of our findings, offering guidance on selecting appropriate short-term demand forecasting models, insights into the application of the predict-then-cluster framework for operational decision-making, and potential directions for future research.}

\subsection{\textcolor{black}{Managerial insight from the short-term demand forecasting models for meal delivery services}}
\label{subsec:managerial_prediction}

\subsubsection{Short-term demand forecasting model selection for on-demand services}

The results of our forecasting experiments highlight ensemble-learning models as the preferred choice for short-term demand prediction in meal delivery platforms, particularly when compared to traditional time series models such as SARIMA and SARIMAX. While traditional models excel in explainability, their inability to handle complex seasonality, such as the joint daily and weekly patterns observed in meal delivery demand, makes them less suitable for this domain. Despite the well explainablility, SARIMA and SARIMAX's model complexity increases drastically with the length of the periodic sequence.
\textcolor{black}{As evidenced by our experiments, SARIMA and SARIMAX failed to complete training within an hour using real-world datasets from the Europe and Taiwan, which contain 21 weeks and 76 days of training data respectively.}

In contrast, ensemble-learning models, in contrast, offer superior computational efficiency, scalability, and accuracy for both point and distributional forecasting.
Showed in our experiments, the ensemble-learning models require much less computational time for model training, hyperparameter tuning, and prediction generation. 
Moreover, ensemble-learning models generally achieve higher accuracy in both point and distributional forecasting compared to the selected benchmarks, given the same amount of training data. 
\textcolor{black}{These advantages make them particularly attractive for operational environments requiring rapid updates and actionable predictions.}

\subsubsection{Enhancing short-term demand forecasting with lagged-dependent features}

Incorporating lagged-dependent variables, which represent recent demand observations, enhances real-time input data by capturing short-term variations. Our point forecasting experiments indicate that these additional features effectively model short-term temporal fluctuations in demand, complementing contextual temporal features (e.g., hour of the day and day of the week) that capture regular seasonal patterns.
The demand forecasting results for European and Taiwanese use cases, as presented in Table \ref{tab:point} and Tabel \ref{tab:point_tw} in Section \ref{subsec:forecasting_result}, show that the lagged-dependent models consistently achieved the highest point forecasting accuracy among the selected forecasting algorithms. Moreover, the additional lagged-dependent features also enhance QRF's distributional forecasting performance according to evaluations with MCRPS as depicted in Table \ref{tab:MCRPS}. By analyzing the standard deviations of residuals compared to those of traditional time series models, we verify the ability of ensemble-learning models to consistently deliver robust forecasting accuracy for both point and distributional forecasting tasks. 

\textcolor{black}{Feature importance analysis underscores the significant role of lagged-dependent variables in improving forecasting performance. In both the European and Taiwanese case studies (Figures \ref{fig:SHAP} and \ref{fig:SHAP_TW}), lagged demand features emerged as the second most influential group of attributes, following the hour-of-the-day attribute. Notably, their importance increased in the Taiwanese case, where contextual information such as weather and holidays was unavailable, demonstrating their critical contribution in scenarios with limited auxiliary data}. \textcolor{black}{From a policy perspective, these findings highlight the value of integrating real-time demand into forecasting pipelines. Incorporating recent demand observations during real-time data collection enhances predictive accuracy, particularly under conditions of limited contextual information. The improved demand forecasts enable platforms to make better decisions during periods of dynamic demand fluctuations, ultimately boosting operational efficiency and system performance.}

\subsubsection{Distributional forecasting performance by QRF predictors and their utilization in dynamic clustering}

\textcolor{black}{In this study, we predict the uncertainty of demand through a distributional forecasting process using QRF. QRF is a non-parametric approach that estimates the predicted distribution of demand for the next 15 minutes by providing quantile predictions. Without pre-assuming a parametric distribution of demand, QRF generates quantile predictions from the subset of historical observations conditioned on the input features, thereby avoiding mismatches between the assumed distribution and actual demand distribution. In our use cases, the quantile predictions are integers since the demand observations are discrete values. Compared to taking quantiles directly from seasonally conditioned historical demand observations, as in the Seasonal benchmark, QRF, and LD-QRF provide a better fit to the underlying distribution on both training and testing data, as shown in Table \ref{tab:MCRPS}.} By estimating demand uncertainty through quantile regression forests, the methodology provides distributional predictions that can improve stochastic optimization, leading to higher robustness in decision-making under uncertain conditions. 

Our predict-then-cluster framework also utilizes quantile predictions to account for demand uncertainty and enhance the predictive clustering process. As analyzed for our European case study with dynamic clustering in Subsection \ref{subsec:clustering_result}, these predicted quantile additions reduce deviations between the clusters generated with predicted and actual demand information.
However, high-quality distributional forecasting requires a sufficient amount of data for training.
Under conditions of limited training data availability, incorporating the quantile predictions from both QRF and LD-QRF may lead to less accurate results compared to the case where \textcolor{black}{only median demand predictions are applied.}

\subsubsection{\textcolor{black}{Utilization of short-term predicted demand to enhance efficiency of real-time operations: demonstration via a fleet rebalancing simulation study}}

Existing literature has extensively addressed operational optimization problems for on-demand mobility and logistics services, such as fleet sizing \cite{xue2021optimization}, fleet reallocation \cite{lei2020efficient}, and order matching \cite{chen2019can}. However, many of these studies either implicitly predict demand within a dynamic optimization framework without evaluating its accuracy or rely on the assumption of constant demand arrival rates. 
\textcolor{black}{Meanwhile, demand forecasting has proven effective in informing forward-looking decisions to improve operational efficiency in real-time policy designs \cite{grahn2021improving}. 
To further demonstrate the practical value of short-term demand predictions in supporting meal delivery platforms, we conducted a simulation study focused on idle fleet rebalancing.}

\textcolor{black}{Based on historical order data from the European use case, this study integrates predictive demand insights into operational planning to evaluate their impact.
Details of the simulation experiments, including the design of the forward-looking idle courier relocation policy, are provided in \ref{appendix:simulation}. Results show that integrating demand predictions significantly enhances platform efficiency by proactive steering of idle couriers toward the areas with higher demand expectations.
When the demand prediction-informed relocation policy is applied, the average order delivery times have been reduced by over 10\% compared to a rule-based policy that does not incorporate predictive insights.}
\textcolor{black}{Beyond the idle fleet rebalancing example, operators of meal delivery and other on-demand services can leverage short-term demand predictions to develop forward-looking policies, enhancing resource allocation efficiency and enabling proactive decision-making driven by predicted demand insights.}

\subsection{Managerial insight from the short-term predict-then-cluster framework}
\label{subsec:managerial_cluster}

The introduced predict-then-cluster framework significantly enhances operational decision-making for on-demand meal delivery and other last-mile mobility and logistics services by dynamically predicting short-term demand and generating a clustered overview of the service network based on future demand. 

Optimization tasks in large urban service networks are often computationally intensive, posing a bottleneck for real-time operations. The predict-then-cluster framework addresses this challenge by clustering areas based on predicted demand and geographical topology, reducing the size of optimization problems and enabling more frequent and efficient decision-making.  \textcolor{black}{Its dynamic clustering approach groups locations into contiguous clusters of areas with similar future demand levels. It creates logical operational units for tasks, streamlining operations such as order batching within clusters or fleet rebalancing across clusters. By reducing the number of geographical units considered in optimization, the application of predict-then-cluster helps lowering computational complexity, enhancing responsiveness, and improving overall operational efficiency.} 

\textcolor{black}{Unlike static clusters derived from historical demand patterns, dynamic clustering adapts to real-time fluctuations in demand, as captured by predictive models. This adaptability ensures that operational strategies remain aligned with evolving demand dynamics, providing actionable and timely spatial insights for decision-making. The ability to adjust clusters based on current predictions fosters a proactive approach to operations, particularly during periods of high variability or peak demand.}

\textcolor{black}{In addition to predictive and operational benefits, the framework offers flexibility and scalability, allowing operators to customize clustering strategies to meet specific needs. For example, clustering outcomes can directly feed into optimization workflows, and the resulting optimization performance can inform adjustments to clustering parameters. This iterative feedback loop enhances the alignment between clustering and optimization, improving overall system performance and enabling adaptive, data-driven decision-making in complex on-demand service environments.}

As introduced in Subsection \ref{subset:formulation}, there are three general types of clustering methods based on the geographical details and predicted demand information of the pick-up zones. 
\textcolor{black}{In the European case study, we analyze the resulting heatmaps using zonal demand estimates without clustering, as well as the heatmaps of clusters generated by thresholding, CKMC, and CCHC-ICE. In the example presented in Figure \ref{fig:heatmap_cluster}, when demand heatmap is generated without clustering or through thresholding}
, the identified high-demand areas appear scattered across the city, reflecting a lack of spatial coherence. These methods are better suited for operations requiring only local comparative insights into demand. \textcolor{black}{For instance, meal delivery platforms can use these outcomes to develop fleet rebalancing strategies by locally prioritizing where to deploy its active couriers among several adjacent pick-up zones based on their comparative future demand level. However, the clusters generated by thresholding do not provide a spatially consistent view of the city, limiting their effectiveness for coordinating city-wide fleet rebalancing.}
Unlike thresholding, CKMC preserves geographical proximity among zones within the same cluster while also considering predicted demand levels. This makes CKMC more suitable for proactive operational strategies where regional consistency is needed. For instance, fleet rebalancing can direct idle couriers toward the direction of high-demand regions in a coordinated fashion. 
Compared to CKMC, \textcolor{black}{CCHC-ICE} ensures that clusters are not only similar in terms of demand but are also geographically contiguous. \textcolor{black}{CCHC-ICE algorithm also offers flexibility in defining user-specified conditions,} making it suitable for cases with specific continuity and cluster configurations.
In real-time management, the dynamically generated geographically contiguous clusters can be utilized as suitable operational units to enhance the optimization scalability of order bundling and delivery routing tasks and improve the overall delivery efficiency. \textcolor{black}{In real-time management, dynamically generated geographically contiguous clusters can be used as operational units to enhance the scalability of optimization tasks such as order bundling and delivery routing, improving overall delivery efficiency. For instance, a delivery route can be suggested to a courier to combine multiple orders within a cluster with high accessibility among its zones.} However, if zones within a cluster are not easily accessible, it can lead to inefficiencies.

In summary, the clustering approach should be carefully selected based on the specific operational requirements. Beyond the real-time operations discussed, the predict-then-cluster framework can also support tactical and macroscopic operations. Tactical decisions, such as fleet sizing, require regular updates, while macroscopic decisions, like facility location planning, are made for the long-term design of the system. Users can adapt the predict-then-cluster framework with predicted demand information updated at different frequencies to generate suitable insights for specific operations. For example, the daily fleet sizing problem can use predicted demand for the next day as input, while location planning decisions can be supported by cluster insights generated with the estimated long-term spatial distribution of demand for a new service region, based on descriptive features of the neighborhoods. Furthermore, operators can tailor their set of features for clustering beyond the predicted demand variables introduced in this study. For example, incorporating the opening hours of restaurants can support crew planning decisions more effectively. Lastly, the short-term predict-then-cluster framework can also generate dynamic clusters based on predicted fleet supply conditions or supply-demand discrepancies in the city. This type of insight can be particularly useful for crowd-sourcing platforms.

\subsection{Future Research Avenues}
\label{subsec:future}

Recent advances in demand forecasting for on-demand services have emphasized methods that capture spatial and temporal dependencies, with promising applications of graph neural networks (GNNs) and convolutional neural networks (CNNs) \cite{rahmani2023graph}. For instance, Chen et al. \cite{chen2022combining} propose a hybrid model combining Random Forest (RF) and Graph WaveNet (GWN), leveraging RF’s variable importance measures to enhance GWN’s prediction accuracy. However, as highlighted by Rahmani et al. \cite{rahmani2023graph}, the sparsity of intermittent demand data presents significant challenges for neural network models. Designed for coarser forecasting granularity, GNNs, and CNNs often under-predict in the presence of intermittent patterns due to their bias toward zero \cite{kourentzes2013intermittent}. In this study, we address this challenge using ensemble-learning approaches, which are more robust to noise and sparsity. Additionally, our demand forecasting models are trained independently for each pick-up zone, limiting their ability to account for spatial dependencies.
Future research could explore neural network models that simultaneously predict demand for all zones while addressing intermittent demand patterns. These approaches should effectively address the challenges posed by intermittent demand patterns. Accuracy may be further improved by extending our proposed ensemble-learning models with spatial contextual information to describe spatial dependency among zones, such as by incorporating the demand correlations between neighboring zones. This enhancement would be particularly beneficial if the distributional forecasting capability could be preserved. 

In Subsection \ref{subsec:managerial_cluster}, we discuss managerial applications for short-term predict-then-cluster framework and analyze the potential applications of different dynamic clustering methods based on geographical requirements within the outcomes.
Future research could explore how real-time operations can benefit from integrating short-term predicted demand and dynamic clustering insights through detailed case studies. 
A promising direction is to extend the predict-then-cluster framework into a dynamic predict-cluster-then-optimize process. This integration would allow for the investigation of whether optimization performance can be enhanced by fine-tuning clustering hyperparameters based on operational outcomes.

\section{Conclusion}
\label{sec:conclusion}
To maintain resilience in service quality, meal delivery platforms must prioritize the timely fulfillment of user orders despite the complexities of stochastic order arrivals and fluctuating demand patterns influenced by factors such as weather, holidays, and time of day. These challenges are compounded by the dynamic movement of couriers, which often leads to imbalances in supply and demand across the service network. Efficient real-time operations are essential for addressing these imbalances, and generating accurate short-term demand predictions is a critical first step. Moreover, the computational complexity of optimizing operations in real time often presents a bottleneck. By clustering service zones based on predicted demand, the complexity of decision-making can be significantly reduced, enabling faster and more efficient operational strategies.

To address these challenges, this study proposes a computationally efficient short-term predict-then-cluster framework, which produces high-quality demand forecasts for each service zone and generates dynamic cluster insights to identify future demand hotspots. 
The demand forecasting component employs lagged-dependent ensemble-learning models to generate accurate point and distributional predictions, incorporating temporal, seasonal, and contextual features. The dynamic clustering component introduces Constrained K-Means Clustering (CKMC) and Contiguity Constrained Hierarchical Clustering \textcolor{black}{with Iterative Constraint Enforcement (CCHC-ICE)} to group service zones based on predicted demand while satisfying geographical and operational constraints.
Experiments are conducted over the empirical order data from a European and a Taiwanese city, where we identified unique weekly, daily, and spatial demand patterns in meal delivery services of these cities.
Forecasting experiments revealed that lagged-dependent ensemble-learning models outperformed traditional benchmarks in both accuracy and computational efficiency across different scenarios. Distributional predictions further improved clustering outcomes by incorporating demand uncertainties, enhancing the robustness of the clustering process. 
\textcolor{black}{In addition, a simulation study is designed to demonstrate the benefits of incorporating short-term demand predictions into proactive real-time operations, such as idle fleet rebalancing for meal delivery services.}

From a managerial perspective, the findings of this study provide valuable guidance for selecting forecasting models tailored to the availability of training data, computational resources, and specific operational requirements. The analysis emphasizes the importance of incorporating temporal and contextual features to enhance short-term forecasting accuracy, effectively capturing periodic demand patterns and recent variations.
The integration of distributional predictions addresses demand uncertainties, offering robust tools to support stochastic optimization in real-time decision-making. By dynamically clustering service zones, the predict-then-cluster framework delivers actionable insights into near-future demand dynamics. 
Furthermore, the proposed framework is adaptable to applications of meal delivery services, as well as other on-demand city logistics and mobility services. Its ability to promote sustainable, efficient, and scalable operations positions it as a versatile solution for urban logistics challenges.

%% References
\bibliographystyle{elsarticle-num}
\bibliography{reference}

\appendix
\section*{Appendices}
\section{\textcolor{black}{Supplementary material for the European use case}}

\subsection{\textcolor{black}{Data anonymization process for synthetic data generation}}
\label{appendix:anonymization}
\textcolor{black}{To ensure the anonymity of the dataset while preserving its utility for reproducible research, a data anonymization process is applied to the original meal delivery demand data to generate a synthetic dataset for the European use case. The process consists of two parts: spatial transformation and demand value standardization. Together, these steps safeguard sensitive information, ensuring that neither spatial nor demand-related details can be used to deduce the identity of the city or the specific operational data of the platform, while still enabling meaningful analysis of the anonymized data.}

\textcolor{black}{Firstly, spatial transformation is performed over the hexagonal zones (defined by the H3 geospatial indexing system) to obscure the exact geographical location of the studied European city. Specifically, the center coordinates of all hexagonal zones were uniformly shifted by a fixed offset. This transformation preserves the relative spatial relationships between zones while ensuring that the original city’s location cannot be identified from the data.}

\textcolor{black}{Secondly, to protect the confidentiality of demand patterns, the order demand values, which are aggregated into 15-minute intervals for each pick-up zone in the forecasting experiments, were standardized. The standardization process involves adjusting the values to have a mean of zero and a standard deviation of one for each pick-up zone. This ensures that temporal demand trends remain comparable across zones, while the absolute demand volumes cannot be traced back to the original data.}

\subsection{\textcolor{black}{Supplementary data analysis}}
\label{appendix:dataset}

\begin{figure}[h!]
    \centering
    \includegraphics[width=0.75\textwidth]{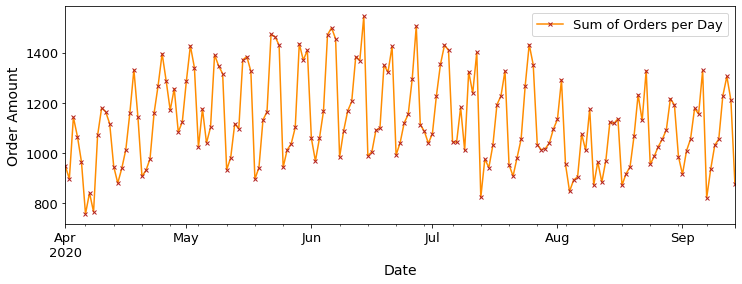}
    \caption{\small The daily total number of orders received in the \textcolor{black}{European} city from the case study.}
    \label{subfig:Dailysum}
\end{figure}

\begin{figure}[h!]
    \centering
    \includegraphics[width=0.75\textwidth]{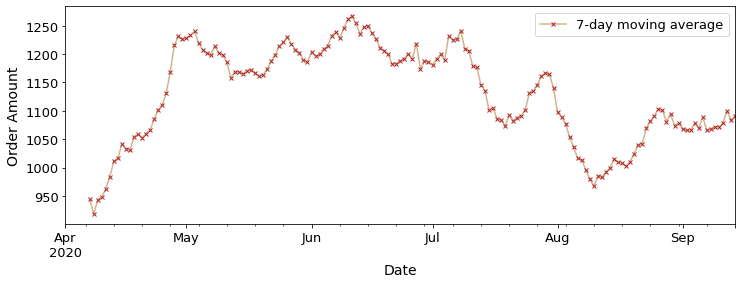}
    \caption{\small The 7-day moving average series of the number of orders received daily in the city.}
    \label{subfig:masum}
\end{figure}

Figure \ref{subfig:Dailysum} shows the daily total number of orders received in this European city from April $1^{st}$, 2020 to September $14^{th}$, 2020. The visualized daily demand time series indicates a recurrent weekly demand pattern. To further inspect potential trends in the time series, we plot the 7-day moving average series of the daily order data with the dates as the X-axis in Figure \ref{subfig:masum}. The moving averages are calculated as
\begin{equation*}
    MA_t = \frac{1}{7} \sum_{i = t-6}^{t} x_{i}, \qquad t = 7,8,\dots,T,
\end{equation*}
where $t$ is the index of the date, which starts from 7 and ends at $T = 166$, the last day covered by the data. 
Figure \ref{subfig:masum} shows a clear upward trend from the beginning of April to the start of May, after which the moving averages stay at a comparatively constant level between May and July before the dive-and-rise patterns between July and September. The moving-average plot suggests a potentially significant trend in the daily order time series. These shifts in demand may be caused by the Covid-19-related measures implemented during the time. 
We cannot yet conclude a yearly seasonality pattern exists for the demand time series since the available data only covers a timeline for less than half of a year. A more extended time series is needed to account for potential annual seasonal patterns in the future.

Based on the analysis of the dual-seasonal pattern \textcolor{black}{discussed in Subsection \ref{subset:cases}}, we can further decompose each business day into 5 periods: breakfast (10:30-11:00), lunch (11:00-14:00), afternoon (14:00-17:00), dinner (17:00-21:00) and night (21:00-21:30). 
During breakfast time, there is an average of 0.14 orders received per 15 minutes with a standard deviation (std) of 0.26. An average of 0.64 (std: 0.73) orders were received during 15-minute intervals for lunch times, 0.68 (std: 0.62) for afternoon times, 2.58 (std: 1.47) for dinner times, and 1.0 (std: 0.86) for night times, respectively.

\section{Formulation and pseudocodes of forecasting algorithms}
\label{appendix:cchc}
\label{appendix:methodology}

\subsection{Random Forest Regression (RF)}
\label{appendix:rf}
The ensemble-learning approach \textcolor{black}{involves} models that
\textcolor{black}{aggregate} the outcomes from a pool of basic models to enhance prediction accuracy and robustness.
Random forest regression/regression forest (RF) is a popular 
ensemble-learning algorithm \textcolor{black}{using} Classification and Regression Tree (CART) as its basic regressor. 
\textcolor{black}{In the previous literature}, regression forest has showed robust and accurate performance \textcolor{black}{for forecasting tasks} and is computationally efficient even with large datasets \cite{speiser2019comparison}.
\textcolor{black}{Initially} proposed by Breiman \cite{breiman2001random}, regression forest simultaneously generates \textcolor{black}{multiple} regression trees, where each tree processes a bootstrap sample of the training data.  
Each tree \textcolor{black}{$h_b$} in the forest \textcolor{black}{$H$ with $B$ trees is generated by iterative node splitting, where data is split at each node to minimize Mean Squared Error (MSE). For each split,} the best feature among the random subset of features $f$ \textcolor{black}{is selected out of} the total set of features $F$. 
\textcolor{black}{This randomness in feature selection reduces the risk of overfitting by} reducing the correlation among trees, \textcolor{black}{thereby} increasing this ensemble model's robustness. \textcolor{black}{Training efficiency and overfitting prevention can also be improved by implementing common stopping criteria, such as maximum tree depth, minimum samples per leaf, minimum impurity reduction, etc.}
\textcolor{black}{In regression forest}, the final result is usually \textcolor{black}{taken as} the average among the results given by all the tree estimators \cite{liaw2002classification}. 

Let us define $y$ \textcolor{black}{as} the \textcolor{black}{target} variable, $X$ \textcolor{black}{as} the \textcolor{black}{predictive} features, and $H := \{h_1,h_2,\dots,h_B\}$ \textcolor{black}{as} the trained regression forest model, where $h$ are the single tree regressor\textcolor{black}{s} of $H$. Assume that there is a new feature vector $X=x$ to be predicted, we can define the prediction of a single tree $h$ by
\begin{equation} \label{eq:singletree}
    \text{single tree: \quad} \hat{\mu}(x,h) = \sum_{i=1}^{n} \omega_i(x,h) \cdot y^{*}_i ,
\end{equation}
where $y^{*}$ is the historical target values assigned to tree $h$ and $\omega_i$ is a weight function which sums to 1 for all $i$. It can be defined as 
\begin{equation} \label{eq:weight_ih}
    \omega_i(x,h) = \frac{\mathds{1}(X_i\in R_{l(x,h)})}{\#\{j: X_j \in R_{l(x,h)}\}},
\end{equation}
where $l(x,h)$ is the leaf
that $x$ ends at tree regressor $h$. The indicator function $\mathds{1}(\cdot)$ equals 1 if $X_i$ shares the same leaf $l(x,h)$ with $x$, otherwise 0. The mathematical expression $\#$ is short for cardinality. And $l(x,h)$ corresponds to a rectangular subspace $R_{l(x,h)}$. 

In regression forest, the point forecasts are produced by averaging the results from \textcolor{black}{all} trees. Hence, the average weight function of $x$ by regression forest $H$ \textcolor{black}{can be rewritten} as,
\begin{equation} \label{eq:weight_rf}
    \omega_i(x) = \frac{1}{B} \sum_{h \in H} \omega_i(x,h).
\end{equation}
Thus, the point prediction of $x$ \textcolor{black}{of regression forest} is given by
\begin{equation} \label{eq:point_rf}
    \text{regression forests:\quad} \hat{\mu}(x) = \sum_{i=1}^{n} \omega_i(x) \cdot y_i,
\end{equation}
which is a weighted sum over \textcolor{black}{target values} $y$ from the training data \cite{meinshausen2006quantile}. The training process of random forest is presented in pseudo-code \ref{alg:rf}.  For the further mathematical details \textcolor{black}{on} RF, we refer to Breiman's \textcolor{black}{and  Meinshausen's literature} \cite{meinshausen2006quantile, breiman2001random}.

\begin{algorithm}
\caption{Regression Forest (RF)} \label{alg:rf}
\begin{algorithmic}[1]
\State \textbf{Input:} Training \textcolor{black}{data}: $S := (X_1, y_1), \dots, (X_n, y_n)$, the set of features: $F$, the number of tree regressors in forest: $B$, and \textcolor{black}{a set of} stopping criteria for tree regressor: $STOP$.
\State \textbf{Output:} A trained random forest model $H$.
\State
\Function{Single Tree Regressor}{$S, F, STOP$}
    \State $h \gets$ initialize the current tree model with $S$ and $F$
    \Repeat{any $STOP$ is activated}
        \State Assign random subset $f$ of $F$ to current node $d$
        \State Split the current data \textcolor{black}{sub}set $S'$ on best feature in $f$
        \State Record the current node $d$ to $h$
        \State Create and move on to the next node
    \Until{any $STOP$ is activated}
    \State \Return $h$
\EndFunction
\State
\Function{Regression Forest}{$S, F, B, STOP$}
    \State $H \gets \emptyset$
    \For{$b \in \{1, \dots, B\}$}
        \State $S_{(b)} \gets$ A bootstrap sample from $S$
        \State $h_{(b)} \gets$ \Call{Single Tree Regressor}{$S_{(b)}, F, STOP$}
        \State $H \gets H \cup h_{(b)}$
    \EndFor
    \State \Return $H$
\EndFunction
\end{algorithmic}
\end{algorithm}

\subsection{eXtreme Gradient Boosting (XGBoost)}
\label{appendix:xgboost}
Gradient boosting is an ensemble machine learning technique that minimizes the loss function by constructing weak learners in a step-wise manner to compensate for the remaining residuals from previous using gradient descent
\cite{friedman2001greedy}, where the series of weak learners are often trees. Therefore, it is also called the gradient-boosted trees method.  
As a regularized variant of gradient boosting, eXtreme Gradient Boosting \textcolor{black}{(XGBoost)}, is a computationally efficient and effective implementation of gradient boosting \cite{chen2016xgboost}. 
Except for the improvements in the computational structure to allow fast training, \textcolor{black}{XGBoost} also prevents overfitting by adopting advanced regularization L1 and L2 \cite{chen2016xgboost}. 
It has achieved state-of-the-art results on a wide range of regression problems and is often seen as the winning solution of many machine learning competitions \textcolor{black}{\cite{bojer2021kaggle}}. 

The training process of XGBoost is listed in pseudo-code \ref{alg:xgb}.

\begin{algorithm}
\caption{\textcolor{black}{eXtreme Gradient Boosting} (XGBoost)} \label{alg:xgb}
\begin{algorithmic}[1]
\State \textbf{Input:} Training dataset $S := (X_1, y_1), \dots, (X_n, y_n)$, the set of features $F$, the number of weak learner tree regressors in forest $T$, and the stopping criteria for tree regressor $STOP$.
\State \textbf{Output:} A trained XGBoost model $H$.
\State

\Function{XGBoost}{$S, F, T, STOP$}
    \State Initialize model: $H_0(x) \gets 0$
    % \State
    \For{$t \in \{1, \dots, T\}$}
        \State Compute negative gradient: $g_t \gets -\nabla L(y, H_{t-1}(x))$
        \State Fit a weak learner to the gradient: $h_t \gets \text{WeakLearner}(X, g_t, F, STOP)$
        \State $h_t \gets$ Regularization($h_t$)
        \State Compute optimal step size: $\gamma_t \gets \arg\min_\gamma \sum_i L(y_i, H_{t-1}(x_i) + \gamma h_t(x_i))$
        \State Update the model: $H_t(x) \gets H_{t-1}(x) + \gamma_t h_t(x)$
    \EndFor
    \State \Return $H$
\EndFunction
\end{algorithmic}
\end{algorithm}

\subsection{Quantile regression forest}
\label{appendix:qrf}
Proposed by Meinshausen \cite{meinshausen2006quantile},
quantile regression forest (QRF) \textcolor{black}{enables distributional forecasting} while reserving the properties of RF. Given the set of observations $Y$ from the training data and the weight function (Equation \eqref{eq:weight_rf}) of random forest, the conditional distribution function of $Y$ can be \textcolor{black}{estimated as the following} \cite{meinshausen2006quantile}:

\begin{equation} \label{eq:qrf_cdf}
    \hat{F}(y|X=x) = E(\mathds{1}(y \geq \hat{Y})) | X=x) = \sum_{i=1}^{n} \omega_i(x) \cdot \mathds{1}(y \geq \hat{Y_i}).
\end{equation}
\textcolor{black}{Hence, for any given quantile $q$ and input vector $x$, QRF generates the quantile prediction $\hat{y}_q(x)$ by following}
\begin{equation} \label{eq:qrf_quantile}
    \hat{y}_q(x) = \inf \left\{ y : \sum_{i=1}^n \omega_i(x) \mathds{1}(y_i \leq y) \geq q \right\}.
\end{equation}

The algorithm of the quantile regression forest method is \textcolor{black}{provided} in pseudo-code \ref{alg:qrf}. Note that the training process is mainly the same as the regression forest, \textcolor{black}{except that} QRF records all the observations on each leaf of every \textcolor{black}{regression} tree \textcolor{black}{to span the distributional predictions, rather than} merely \textcolor{black}{taking} the average. 

\begin{algorithm}
\caption{Quantile Regression Forest \textcolor{black}{(QRF)}} \label{alg:qrf}
\begin{algorithmic}[1]
\State \textbf{Input:} \textcolor{black}{Training data $\mathcal{S}$}, New feature vector $x$
\State \textbf{Output:} \textcolor{black}{Quantile predictions for $x$}
\State
\Function{\textcolor{black}{Quantile Regression Forest - Training}}{$S, F, B, STOP$}
    \State Train a random regression forest $H$ following \Call{Regression Forest}{$S, F, B, STOP$}
    \State Record all training samples in each leaf of $H$
    
    \State \Return $H$
\EndFunction
\State
\Function{\textcolor{black}{Quantile Regression Forest - Forecasting}}{$H$, $x$, $q$}
    \For{tree $h \in H$}
        \State Run $h$ with $x$;
        \State Calculate weight $\omega_i(x, h)$ for each training sample $i$, following Eq. \eqref{eq:weight_ih}.
    \EndFor
    % \State
    \State Calculate average weight $\omega_i(x)$ for each $i$ across all trees, following Eq.\eqref{eq:weight_rf}
    
    \State Generate quantile \textcolor{black}{predictions $\hat{y}_q(x)$} using $\omega_i(x)$ for quantile $q$, \textcolor{black}{following Eq.\eqref{eq:qrf_quantile}}
    \State \Return $\hat{y}_q(x)$
\EndFunction

\end{algorithmic}
\end{algorithm}

\subsection{Deterministic forecasting benchmarks: TBATS, SARIMA, and SARIMAX}
 \label{appendix:point_benchmarks}
Advanced time series predicting technique TBATS is a modified state-space model for exponential smoothing method invented by De Livera et al. \cite{de2011forecasting}. TBATS incorporates Fourier terms, Trigonometric seasonality, Box-Cox transformation, ARMA error correlations, and Trend and Seasonal components. 
TBATS is not only capable of handling time series with complex seasonality but also selecting the time series parameters in a completely automatic manner using the Akaike information criterion (AIC) \cite{hyndman2018forecasting}. This makes TBATS comparable to machine learning approaches in terms of \textcolor{black}{simplicity of application}, while the other traditional time series models like ARIMA often require professional knowledge for hyperparameter selection. 
Therefore, in this study, univariate TBATS is implemented as our state-space benchmark model.  For mathematical details of the TBATS model, we refer to the literature \cite{de2011forecasting}. 

Traditional time series prediction models SARIMA, and SARIMAX have been taken as benchmark models in previous demand forecasting literature for on-demand meal delivery services \cite{hess2021real,yu2023short}.
The regression of ARIMA consists of three major components: autoregressive (AR) terms that capture relationships between the current and past observations, differencing (I) term that ensures stationarity, moving average (MA) terms that estimate the dependencies between historical prediction errors and the current observation. SARIMA extends ARIMA by capturing seasonality effects. SARIMAX is the extension of SARIMA that includes exogenous variables for predictions.
However, it may be challenging to calibrate SARIMA and SARIMAX models with a long seasonal sequence. For such cases, the computational cost of hyperparameter tuning can be significantly higher if the modeling process should be fully automated. Alternatively, experts can be hired to formulate these models through a series of statistical tests. \textcolor{black}{Despite these challenges}, comparing the point prediction results of SARIMA and SARIMAX with the selected ensemble-earning predictors provides valuable insights into understanding the trade-off between computational cost and the accuracy of demand predictions.

\section{Candidate hyperparameters values for \textcolor{black}{demand forecasting} model tuning} \label{appendix:hyperparam}
The hyperparameter values we try for the model tuning of ensemble-learning predictors are listed in Table \ref{tab:hyper_param}.
\begin{table}[h!]
    \centering
    \caption{The candidate hyperparameter values for demand forecasting ensemble-learning models.}
    \resizebox{0.68 \textwidth}{!}{%
    \begin{tabular}{ll|ll}
    \hline
    \hline
    \multicolumn{2}{c}{\textbf{RF \& LD-RF \& QRF \& LD-QRF}} & 
    \multicolumn{2}{c}{\textbf{XGBoost \& LD-XGBoost}} \\
    \hline
    \textit{n\_estimators} & 50, 75, \dots, 175, 200 & \textit{n\_estimators} & 50, 75, \dots, 175, 200\\
    \textit{max\_features} & `auto', `sqrt' & \textit{learning\_rate} & 0.1, 0.15, 0.2, 0.25, 0.3\\
    \textit{max\_depth} & 3,4,5,6,7 & \textit{max\_depth} & 3,4,5,6,7\\
    \textit{min\_samples\_split} & 4,6,8,10 & \textit{subsample} & 0.5, 0.75, 1.0\\
    \textit{min\_samples\_leaf} & 2,3,4,5,10 & \\
    \hline \hline
    \multicolumn{4}{l}{\footnotesize Note: We refer to the documentation of model packages by scikit-learn \cite{scikit-learn} for an elaborate explanation} \\
    \multicolumn{4}{l}{\footnotesize of the hyperparameters.}
    \end{tabular}}
\label{tab:hyper_param}
\end{table}

\section{\textcolor{black}{Python packages used for model implementation}}
For model implementation, we use package pmdarima \cite{pmdarima} for SARIMA and SARIMAX models, a package \cite{package_TBATS} written by Skorupa for TBATS model, package scikit-learn \cite{scikit-learn} for RF and XGBoost model, package quantile-forest \cite{package_QRF} for QRF model. 

\section{\textcolor{black}{Pseudocode implementation of the CCHC-ICE framework with example user-specified constraints}}
\textcolor{black}{This appendix presents the pseudocode for the Contiguity Constrained Hierarchical Clustering with Iterative Constraint Enforcement (CCHC-ICE) framework introduced in Section \ref{sect:CCHC-ICE}. Three kinds of user-specified constraints are introduced in this example implementation: the cluster dissimilarity threshold, minimum number of clusters, and maximum cluster size. Algorithm \ref{alg:CCHC} outlines the main iterative clustering process, explaining the inputs and outputs of the framework. Algorithm \ref{alg:CCHC_helper} provides the helper functions necessary for enforcing geographical contiguity and user-specified constraints, updating adjacency relationships, and managing cluster operations. Together, these pseudocodes detail the implementation of the CCHC-ICE framework, demonstrating how constraints are dynamically enforced during the clustering process, as applied in the European case study. This serves as a guide for implementation and replication.}

\begin{algorithm}
\caption{Contiguity Constrained Hierarchical Clustering \textcolor{black}{with Iterative Constraint Enforcement (CCHC-ICE)}} \label{alg:CCHC}
\begin{algorithmic}[1]
\State \textbf{Standard Input:}
\begin{itemize}
    \item \textbf{Predicted Demand Features:} $\mathbf{X}_t$, an $N \times M$ matrix, where $N$ is the number of zones and $M$ the length of predicted demand inputs, at current time step $t$;
    \item \textbf{Zone-wise Adjacency Matrix:} $\mathbf{C}_z$, geographical adjacency among zones;
    \item \textbf{Feature Importance Vector*:} $w$,  
    % an $M \times 1$ weight vector for the calculation of similarity matrix (only required when $M \geq 2$).
    an $M \times 1$ weight vector where $w_j$ correspond to the feature importance for feature $j$ for the calculation of similarity matrix (only when $M \geq 2$).
    \end{itemize}
\State \textbf{\textcolor{black}{User Specified Constraints Input:}}
\begin{itemize}
    \item \textbf{Minimum Number of Clusters:} $K_{\text{min}}$, a hyperparameter which indicates the minimum number of clusters. Hierarchical clustering stops when $K \leq K_{\text{min}}$;
    \item \textbf{Maximum Size per Cluster :} $s_{\text{max}}$, a hyperparameter that controls the maximum number of pickup zones that can be clustered together;
    \item \textbf{Cluster-distance Threshold :} $D_{\text{max}}$, the maximum distance value between clusters to be merged.
\end{itemize}

\State \textbf{Output:}
\begin{itemize}
    \item \textbf{Cluster labels:} $L$, an $N \times 1$ vector recording the final labels for all zones;
    % corresponding to the resulting cluster label of each pickup zone;
    \item \textbf{Dictionary of clusters:} $\Omega = \{\omega_1, \omega_2, \cdots, \omega_{K}\}$, where each cluster element $\omega$ contains all the pickup zones belonging to this cluster, i.e., with label $l$.
\end{itemize}

\State

\Function{Initialize helper variables}{$N$}:
\State $\Omega \leftarrow \{\omega_1, \omega_2, \cdots, \omega_{N}\}$, where each pickup zone is an individual cluster initially.
\State $K \leftarrow N$, the initial number of cluster is the number of zones.
\State Set \textit{contiguity violation} $\leftarrow$ False
\State \Return $\Omega$, $K$, \textit{contiguity violation}
\EndFunction
\State

\Function{\textcolor{black}{CCHC-ICE}}{$\mathbf{X}_t, \mathbf{C}_z, K_{\text{min}}, s_{\text{max}}$}
\State $\Omega$, $K$, \textit{contiguity violation} $\gets$ \Call{Initialize helper variables}{$N$}
\While{$K \geq K_{\text{min}}$ and \textit{contiguity violation} = False}
    \State $\mathbf{D}_{s} \gets$ \Call{Constrained Distance Matrix Update}{$\mathbf{C}_z, \Omega, \mathbf{X}_t$}
    \State $\Omega' \gets$ \Call{Agglomerative Hierarchical Clustering}($\mathbf{D}_{s}, K$, $D_{\text{max}}$)
    \State $\text{contiguity violation} \gets$ \Call{check-contiguity-conditions}($\Omega', \mathbf{C}_e, \mathbf{C}_p$)
    \If{\textit{contiguity violation} = True \textbf{or} $\Omega = \Omega' $}
        \State \textbf{break}
    \Else
        \State $\Omega \gets \Omega' $ 
        \State $K \gets K - 1$
    \EndIf
\EndWhile
\State \Return $\Omega$
\EndFunction

\end{algorithmic}
\end{algorithm}

\begin{algorithm}
\caption{Helper functions for \textcolor{black}{CCHC-ICE}} \label{alg:CCHC_helper}
\begin{algorithmic}[1]
\Function{Within Cluster Connectivity}{$\Omega$}:
\State Initialize zone connectivity matrix: $\mathbf{C}_e \gets$ zero-matrix of shape $N \times N$ 
\For{$i \gets 0$ \textbf{to} $N$}
    \For{$j \gets 0$ \textbf{to} $N$}
    \If{$i \neq j$ and zone $i$ and zone $j$ in the same cluster according to $\Omega$}
        \State $\mathbf{C}_e [i,j] \gets 1$
    \Else
        \State $\mathbf{C}_e [i,j] \gets 0$
    \EndIf
    \EndFor
\EndFor
\State \Return $\mathbf{C}_e$
\EndFunction
\State

\Function{Cluster Adjacency Constrained Connectivity}{$\mathbf{C}_z$, $\Omega$, $s_{max}$}:
\State Initialize zone connectivity matrix: $\mathbf{C}_p \gets$ zero-matrix of shape $N \times N$
\State $\Psi = [\psi_1, \cdots, \psi_{\omega}]$ where $\psi_w$ is the list of feasible contiguous clusters for a cluster $c_w$, based on the zone-wise adjacency matrix $\mathbf{C}_z$ and maximum cluster size constraint $s_{max}$.
\For{i in 0 to N}
	$c_w \gets$ current cluster of zone $i$ 
	\For{j in 0 to N}
		\If{$i \neq j$ and zone $j$ belongs to a contiguous cluster defined in $\psi_w$}
			\State $\mathbf{C}_p [i,j] \gets 1$
		\Else
			\State $\mathbf{C}_p [i,j] \gets 0$
		\EndIf
	\EndFor
\EndFor
\State \Return $\mathbf{C}_p$
\EndFunction
\State

\Function{Constrained Distance Matrix Update}{$\mathbf{C}_z, \Omega, \mathbf{X}_t$}:
    \State Initialize contiguity constrained distance matrix: $\mathbf{D}_z \gets$ $N \times N$ zero-matrix 
    \State $\mathbf{C}_e$(Existing connection matrix) $\gets$ \Call{Within Cluster Connectivity}{$\Omega$} 
    \State $\mathbf{C}_p$(Potential connection matrix) $\gets$ \Call{Cluster Adjacency Constrained Connectivity}{$\mathbf{C}_z$, $\Omega$}
    \For{i in 0 to N}
	\For{j in 0 to N}
		\If{$\mathbf{C}_e [i,j] = 1$}
  \State \# if connection is reserved since two zones already in the one cluster,
			\State $\mathbf{D}_z [i,j] \gets 0$
		\ElsIf{$\mathbf{C}_p [i,j] = 1$}
  \State \# if connection is possible since they belong to two adjacent clusters, 
			\State $\mathbf{D}_z [i,j] \gets \textit{distance}(\mathbf{X}_t[i,:], \mathbf{X}_t[j,:])$, according to Eq.\eqref{eq:clustering_distance}. 
            \Else
\State \# if the connection doesn't satisfy contiguity constraints,
                \State $\mathbf{D}_z [i,j] \gets +\infty$
		\EndIf
	\EndFor
\EndFor
    
\State \Return $\mathbf{D}_z$
\EndFunction
\State
\end{algorithmic}
\end{algorithm}

\section{Evaluation metrics formulation}
\label{subset:metrics}

\subsection{Point Forecasting Metric: MAE, RMSE, RMSLE}

Given the total number of prediction equals to $T$ for each pick-up zone $i$, the actual order demand vector $y^i$, and corresponding predicted demand vector $\hat{y}^i$,
the mean absolute error for pick-up zone $i$ is calculated as,
\begin{equation}
    \text{MAE}^{i} = \frac{\sum^{T}_{t=1} |y^{i}_t - \hat{y}^{i}_t|}{T} .
\end{equation}
The root mean squared error is calculated as, 
\begin{equation}
    \text{RMSE}^{i} = \sqrt{\frac{\sum^{T}_{t=1}(y^{i}_t - \hat{y}^{i}_t)^2}{T} }.
\end{equation}

Although Mean Absolute Percentage Error (MAPE) is a common metric for other demand forecasting
problems, it is not suitable for our case. Our data analysis shows that it is common for many pick-up zones to receive
no orders during several 15-minute time windows during the off-peak hours, which means there are many
zero values on the demand time series. Hence, using MAPE directly will encounter the divided-by-zero
problem. 
We utilize an alternative approach, Root Mean Square Logarithmic Error (RMSLE), as a relative error measure that does not suffer the zero-demand problem. It is formulated as:
\begin{equation}
    \text{RMSLE}^{i} = \sqrt{\frac{\sum^{T}_{t=1}(log(\hat{y}^{i}_t + 1) - log(y^{i}_t+1))^2}{T} }.
\end{equation}
RMSLE is a relative error measure that does not penalize large residuals when both predicted and actual values are large. Moreover, it penalizes underestimates more than overestimates. This property is suitable for cases where having extra inventory or supply is preferable to failing to meet the demand.
This approach has also been applied to other demand forecasting studies of on-demand shared mobility services \cite{qiao2021dynamic, jimenez2021can}.

\subsection{Distributional Forecasting Metric: MCRPS}

CRPS quantifies the discrepancy between the predicted cumulative distribution function (CDF) resulting from the forecast and the `actual' CDF. The `actual' empirical CDF of the scalar observation $y$ is simply represented as a step function $\mathbf{1}(\hat{y} \geq y)$ that returns 1 where $x$ is greater or equal to the observation and returns 0 otherwise. Following the instantaneous form formulation in Gneiting and Raftery \cite{gneiting2007strictly}, CRPS is defined as, 
\begin{equation}
    \mathrm{CRPS}(F, y) = \int_{\mathbb{R}} \left[ F(\hat{y}) - \mathbf{1}(\hat{y} \ge y) \right]^2 \, d\hat{y}.
\end{equation}
\textcolor{black}{By taking the integral over the squared difference between the target and estimated distributions, CRPS values are non-negative, ranging from 0 to $+ \infty$.}
\textcolor{black}{In our study, nonparametric forecasting method QRF only provides quantile predictions for a given quantile value $q_\kappa \in [0,1]$, instead of a continuous distribution. Therefore, an empirical CDF $\hat{F}(\hat{y})$ needs to be estimated from a collection of quantile predictions $Q(q_\kappa)$ corresponds to quantiles $q_1, q_2, \cdots, q_K$, } 

\begin{equation}
    \hat{F}(\hat{y}) =
    \begin{cases} 
    0, & \text{if } \hat{y} \le Q(q_1), \\
    q_\kappa + \frac{(\hat{y} - Q(q_\kappa))}{(Q(q_{\kappa +1}) - Q(q_\kappa))} \cdot (q_{\kappa +1} - q_\kappa), & \text{if } Q(q_\kappa) \le x \le Q(q_{\kappa +1}), \\
    1, & \text{if } \hat{y} \ge Q(q_K).
    \end{cases}
\end{equation}

\begin{figure}
    \centering
    \includegraphics[width=0.6\textwidth]{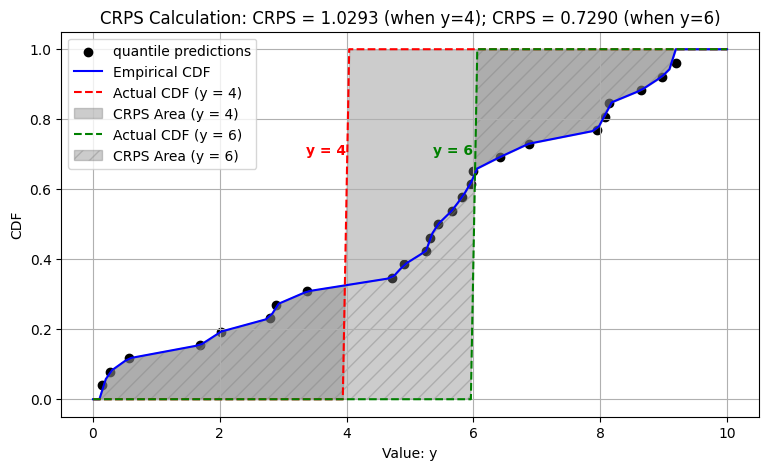}
    \caption{Visualized demonstration of CRPS calculation for a group of 25 sample quantile predictions (corresponding to quantile values $q = 0.04, 0.08, \cdots 0.96$ ) and two different target values, $y=4$ and $y=6$.}
    \label{fig:CRPS_example}
\end{figure}

\textcolor{black}{Figure \ref{fig:CRPS_example} demonstrates how CRPS is estimated by measuring the discrepancy between the estimated empirical CDF, which is fitted from a group of quantile predictions $Q(q_\kappa)$, and the actual empirical CDF generated by a step function from a scalar observation $y$. 
In this study, we select $K=9$ quantile predictions corresponding to quantile values $q = 0.1, 0.2, \cdots, 0.9$ to calculate $\text{CRPS}^{i}_t (\hat{F_t}, y_t)$ each pick-up zone $i$ for time interval $t$. 
We only consider a limited number of quantiles for the evaluation because both QRF and LD-QRF generate quantile predictions from historical observations, while the historical demand values typically fall within a narrow range for most time windows. 
The mean CRPS (MCPRS) for zone $i$ will be computed as the average among all the forecasting time steps, }

\begin{align}
        MCRPS^{i} &= \frac{\sum^T_{t=1} \text{CRPS}^{i}_t (\hat{F_t}, y_t)}{T} 
\end{align}
A lower MCRPS value indicates that the quantile predictions are closer to the underlying actual demand distribution in general, therefore suggesting a better distributional forecasting performance by the model.
\section{Simulation study on utilizing short-term demand forecasting to enhance meal delivery system efficiency}
\label{appendix:simulation}
\textcolor{black}{In line with our discussion on the potential implementation of the proposed demand forecasting model and its policy-making benefits, we demonstrate how high-quality short-term demand predictions can support real-time operations of on-demand meal delivery services through a simulation experiment.
Specifically, we introduce an idle fleet rebalancing strategy to relocate idle couriers towards under-supplied regions in the future based on short-term demand predictions of the service network.
As an extension of the European case study, a simulation study is designed for the city and performed with the historical order data as input.
The goal is to assess how the overall delivery efficiency is impacted by the implementation of this prediction-informed policy. 
The remainder of this section is organized as follows:  \ref{subsect:simulator} describes the design of the meal delivery fleet simulator, \ref{subsect:relocation} defines the forward-looking idle courier relocation policy, and \ref{subsect:simulation_results} presents the detailed specification of simulations and analyzes the results.}

\subsection{\textcolor{black}{Design of the meal delivery simulations}}
\label{subsect:simulator}
\textcolor{black}{In this section, we describe the design of a simulator that models the movement of an active fleet within a meal delivery service network over the course of a day. The simulator dynamically assigns delivery and relocation tasks to couriers while tracking orders, tasks, and courier statuses. Performance indicators are embedded to assess the impact of different real-time idle fleet rebalancing strategies on overall delivery efficiency.}

\textcolor{black}{The meal delivery service network of a city is represented by a set of individual zones in the simulation, denoted as $Z: {z_1, z_2, \dots, z_{M}}$. As described in Section \ref{subset:formulation}, each meal delivery order is associated with a pick-up zone and a destination zone, representing the restaurant and household locations for the order.}

\textcolor{black}{The simulator begins by initializing a fleet of $J$ couriers ($C: {c_1, c_2, \dots, c_J}$). Couriers are categorized as either idle or busy. An idle courier is available for a delivery or relocation assignment, while a busy courier is engaged in an ongoing task. Upon completing a task, a courier becomes idle at the destination location and is available for a new assignment. Throughout the simulation, the number of idle couriers in each zone varies.}

\textcolor{black}{Orders arrive in the system dynamically during the simulation. An order $o$ only arrives in the system upon arrival time $t^o_a$, and is unknown to the meal delivery platform beforehand. The pick-up and drop-off zone locations of the order are denoted as $z^o_p$ and $z^o_d$. Orders are assumed to be ready for pick-up from the restaurant as soon as they arrive in the platform.}

\textcolor{black}{At each time step, the simulation updates its dynamics. 
Firstly, the simulator goes through the assignment status of busy couriers, releases who just completed the tasks at this time step, and updates them to be idle.
Next, the simulator checks whether one or more orders have arrived to the system. The simulator then checks for new orders, assigning each to the closest available courier from the order's pick-up location. Upon assignment, the delivery time of this task is registered as the sum of the pick-up travel time from this courier's current location to the pick-up zone, delivery travel time from the pick-up zone to the destination zone, plus the service time for picking up and delivering the order. If no courier is available for assignment, the order is rejected. Lastly, the system looks for couriers who have stayed idle for a consecutive period beyond a predefined threshold.
These couriers may be assigned to a relocation task following the suggestions from the idle fleet rebalancing algorithm.}

\subsection{\textcolor{black}{Idle courier relocation policies}}
\label{subsect:relocation}
\textcolor{black}{Relocating idle couriers is essential for maintaining an efficient fleet, minimizing downtime, and balancing supply and demand across urban areas. A classic approach used by meal delivery platforms involves relocating idle couriers to the nearest pick-up zone with restaurants. However, this method often ignores the overall supply and demand dynamics of the network, which can lead to fleet imbalances across the city and reduced efficiency.}

\textcolor{black}{With short-term demand predictions, a forward-looking relocation policy can be implemented to improve fleet efficiency. This policy takes into account both idle courier supply and short-term demand forecasts to anticipate supply-demand imbalances across the network. And it addresses them by proactively moving idle couriers to areas expected to have higher demand.} 

\textcolor{black}{
The forward-looking relocation policy contains two parts: local supply-demand imbalance inspection and relocation assignment.
The anticipated future supply-demand deficit $\hat{\Delta}(z)$ is calculated as $\hat{\Delta}(z) = s(z) - \hat{d}(z)$, where $s(z)$ is the current supply of idle couriers at zone $z$ and $\hat{d}(z)$ its predicted demand for the next time interval. For an idle courier ready for relocation, $\hat{\Delta}(z)$ is firstly computed for the candidate zones $Z$, which include the courier's current zone and its neighboring adjacent zones. The zone from $Z$ with the highest estimated future supply-demand deficit is selected. If this is the courier's current zone, they remain in place. Otherwise, the courier is assigned to relocate to the selected zone.}

\subsection{\textcolor{black}{Specifications and results of simulation experiment}}
\label{subsect:simulation_results}

\textcolor{black}{
The meal delivery fleet simulations are specified for the European case study using its actual order transaction data from each workday within the testing data, corresponding to 7 different simulation scenarios.
The simulation adopts the service network topology from the European use case, consisting of a total of 50 zones. Of these, 20 zones can serve as pick-up zones, while all zones can function as destination zones.
The simulation updates at each time step, corresponding to one minute of the day. The simulation's start and end times match the operating hours of the meal delivery service in the European use case. Orders arrive within the simulation according to their actual recorded times in the historical transaction data, ensuring realistic demand dynamics. Couriers are assumed to be active for the entire duration of each simulated day. To simplify the design, the travel times between any two zones in the network are calculated based on the distance derived from a grid traversal function provided by the geo-indexing package H3 for the hexagonal zones defined in the use case \cite{uberh3}. The travel time between adjacent zones is computed using the distance between the centers of the adjacent grids and the local average biking speed. The service times for pick-up at restaurants and delivery at households are assumed to be both 3 minutes, for each order.
The fleet size $J$ is fixed at 30 couriers for each simulation, where the courier is initialized at a randomly selected zone as their starting location.
Lastly, the courier's idle duration threshold for suggesting relocation is set to 5 minutes.}

\textcolor{black}{To assess the impact of the \textit{forward-looking} idle courier relocation policy on system performance, we include two benchmarks: \textit{nearest pick-up zone} and \textit{forward-looking (with actual demand)} idle courier relocation policies. 
The first benchmark is the myopic policy that relocates idle couriers to the nearest pick-up zone. By comparing to this benchmark, we aim to compare the effectiveness of incorporating predictive insights into fleet rebalancing.
The second benchmark replaces the predicted demand $\hat{d}(z)$ by the actual future demand $d(z)$ in the anticipated supply-demand deficit $\hat{\Delta}(z)$ calculations. The inclusion of this benchmark helps to evaluate how forecasting errors affect the rebalancing efficiency.}

\begin{figure}[h!]
    \centering
    \includegraphics[width=0.7\linewidth]{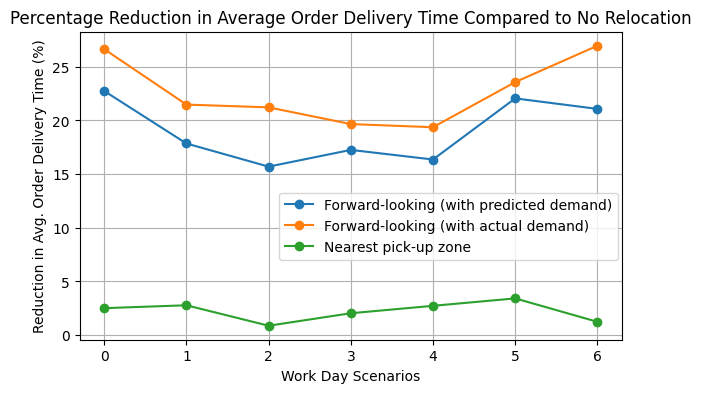}
    \caption{\textcolor{black}{Impact of different relocation policies on the reduction of average order delivery time. The figure shows the percentage reductions in the average delivery time per order compared to no relocation for three policies: forward-looking idle courier relocation with predicted demand, forward-looking relocation with actual demand, and nearest pick-up zone relocation. Results are shown for seven different workday scenarios using historical order data from the European case study, with 100 simulations conducted for each scenario.}}
    \label{fig:simulation_results}
\end{figure}

\textcolor{black}{
Delivery speed is a key performance indicator for on-demand meal delivery services. Short delivery times are crucial for maintaining customer satisfaction. We conducted simulations for each workday scenario to evaluate the effectiveness of three different idle courier relocation policies in reducing average order delivery times, compared to no relocation. Each policy was tested with 100 simulations per scenario to ensure consistency and reliability of results. Figure \ref{fig:simulation_results} shows that both forward-looking policies consistently achieved significantly greater reductions in average delivery time per order compared to the nearest pick-up zone benchmark. The forward-looking policy with demand predictions performed almost as well as the policy using actual demand. These results highlight that accurate demand anticipation is crucial for minimizing delivery times, with even predicted demand providing significant efficiency gains. In contrast, the traditional heuristic approach of nearest pick-up zone relocation led to only minor improvements. In addition, the simulation outcomes suggest that average order rejection rates remained consistent regardless of the relocation policy applied, given the same fleet size.
The simulation results underscore the value of proactive and forward-looking policy-making with short-term demand predictions for on-demand meal delivery services.
}

\end{document}